\documentclass[        pra,
                                twocolumn,
                                groupedaddress,
                                showpacs,
                                ]{revtex4}
\usepackage[T1]{fontenc}      
\usepackage[latin1]{inputenc} 
\usepackage{graphicx}         
\usepackage{setspace}         
\usepackage{bm}                              
\usepackage[usenames,dvipsnames]{color}                        
\usepackage{framed}
\usepackage{placeins} 
\usepackage{amsmath}
\usepackage{amssymb}
\usepackage{dsfont}
\usepackage{bm}
\usepackage{amsmath}
\usepackage{amsfonts}
\usepackage{amssymb}
\usepackage{color}
\usepackage{dsfont}

\newcommand{\ket}[1]{|{#1}\rangle}
\newcommand{\bra}[1]{\langle{#1}|}

\newcommand{\ketbra}[2]{|{#1}\rangle\langle{#2}|}

\newcommand{\beq}{\begin{equation}}
\newcommand{\eeq}{\end{equation}}

\newcommand{\OPrho}{\hat{\rho}}
\newcommand{\OPham}{\hat{\mathcal H}}
\newcommand{\OPp}{\hat{p}}
\newcommand{\OPx}{\hat{x}}
\newcommand{\OPa}{\hat{a}}
\newcommand{\OPadagger}{\hat{a}^{\dagger}}
\newcommand{\OPsigp}{\hat{\sigma}^{\dagger}}
\newcommand{\OPsigm}{\hat{\sigma}}
\newcommand{\OP}[1]{\hat{#1}}
\newcommand{\commutator}[2]{[#1,#2]}
\newcommand{\mean}[1]{\langle #1\rangle}

\begin{document}
\title{Cooling of atomic ensembles in optical cavities: \\Semiclassical limit}
\author{Stefan \surname{Sch\"utz}$^{1}$}
\email[e-mail: ]{stefan.schuetz@physik.uni-saarland.de}
\author{Hessam \surname{Habibian}$^{1,2,3}$}
\author{Giovanna Morigi$^{1}$}
\affiliation{
$^1$ Theoretische Physik, Universit\"at des Saarlandes, D-66123 Saarbr\"ucken, Germany\\
$^2$ Departament de F\'isica, Universitat Aut\`onoma de Barcelona, E-08193 Bellaterra, Spain\\
$^3$ ICFO -- Institut de Ci\`encies Fot\`oniques, Mediterranean Technology Park, E-08860 Castelldefels (Barcelona), Spain
}
\date{\today}

\begin{abstract}
The semiclassical dynamics of atoms are theoretically studied, when the atoms are confined inside a standing-wave high-finesse resonator. The atoms are cooled by scattering processes in which the photons of a transverse laser are coherently scattered into the cavity mode. We derive a Fokker-Planck equation for the atomic center-of-mass variables which allows us to determine the equations of motion in the semiclassical limit for any value of the intensity of the laser field. We extract its prediction for the dynamics when the resonator is essentially in the vacuum state and the atoms are cooled by scattering photons into the cavity mode, which then decays. Its predictions for the stationary atomic distribution are compared with the ones of the Fokker-Planck equation in [P. Domokos, P. Horak, and H. Ritsch, J. Phys. B {\bf 34}, 187 (2001)], which has been derived under different assumptions. We find full agreement in the considered parameter regime.
\end{abstract}

\maketitle
\section{Introduction}
\label{introduction}

The possibility of cooling and trapping the atomic motion by means of electromagnetic radiation has brought to a remarkable advance in atomic physics and quantum optics, which was officially recognized with the Nobel Prize in physics in 1997 \cite{RMP:1997}. In a nutshell, radiative scattering can cool the motion of atoms by means of the mechanical effects of atom-photon interactions. This is achieved by scattering processes, in which the transition rate to states with lower mechanical energy is enhanced by suitably driving an atomic transition, so that the frequency of the absorbed photon is in average smaller than that of the emitted one \cite{Wineland:1978,Stenholm:1986}. In presence of high-finesse optical resonators, these processes can be tailored using the strong coupling with the cavity field \cite{Horak:1997,Vuletic:2000,Domokos:2003,Keever:2003,Maunz:2004,Murr:2006}. 

\begin{figure}[hbt]
\begin{center}
\includegraphics[width=0.4\textwidth]
{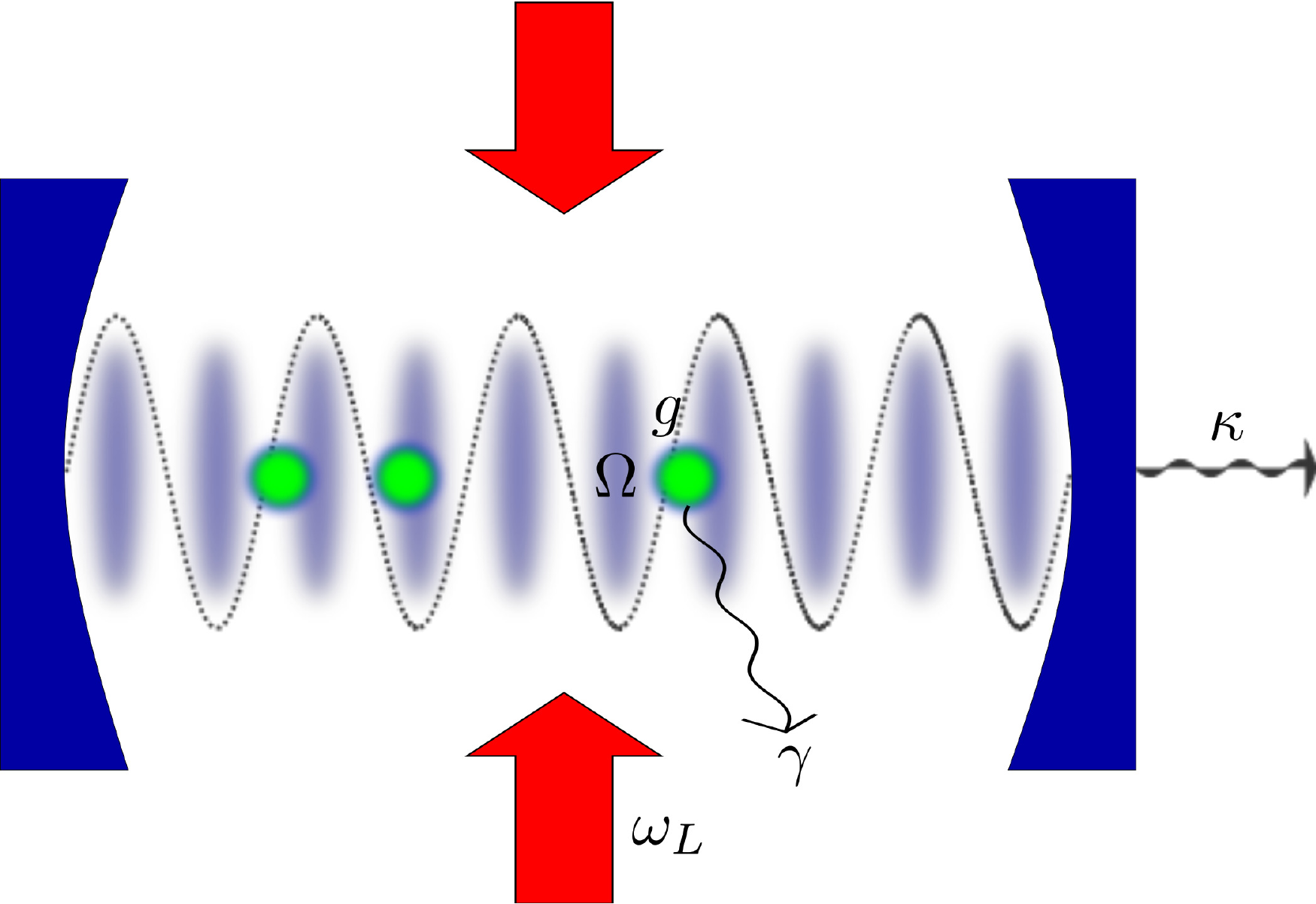}
\caption{\label{Fig:sys} (Color online) A gas of atoms is confined within a standing-wave resonator and is illuminated by a transverse laser. A dipolar transition of the atoms couples quasi-resonantly with the fields and scatters photons from the laser into the cavity. We analyse the dynamics of the atomic center-of-mass motion and steady state in the semiclassical regime.}
\end{center}
\end{figure}

A remarkable property of the mechanical effects of light inside of a high-finesse resonator are the collective phenomena due to multiple scattering of photons, which mediate an effective interaction between the atoms. They give rise to nonlinear dynamics, such as bistability induced by the nonlinear coupling with the motional degrees of freedom \cite{Bistability}, syncronization \cite{Kuramoto}, and collective-atomic recoil lasing \cite{CARL}. In single-mode standing wave cavities they can lead to the formation of spatially-ordered structures \cite{Domokos:2002,Vuletic:2003,Asboth:2005,Baumann:2010,Barrett:2012}. This phenomenon is found in a setup, like the one sketched in Fig. \ref{Fig:sys}, where the atoms are confined inside a resonator and are driven by a transverse laser. It exhibits a threshold, which is mainly determined by the intensity of the laser. Above threshold, ordered atomic structures (Bragg gratings) form which coherently scatter photons into the cavity resonator, and vice versa, the cavity field stably traps the atoms in the grating \cite{Asboth:2005,Vuletic:2003,Baumann:2010}. 

The theory of selforganization in laser-cooled atomic ensembles coupled to cavities has been pioneered by Ritsch, Domokos, and coworkers \cite{Domokos:2013}, who derived a Fokker-Planck equation describing the coupled dynamics of cavity field and atoms in the limit in which the atomic and field variables can be treated semiclassically \cite{Horak:JPB}. On this basis, the selforganization threshold has been determined \cite{Asboth:2005} and numerical simulations of the system dynamics were performed \cite{Domokos:2013}. This theoretical model does not describe, however, the properties of the cavity field, which is treated in the semiclassical limit. The semiclassical  approximation in fact breaks down when the intracavity field is small, namely, below and close to the selforganization threshold. Close to threshold, when the patterns are formed, in particular, fluctuations are expected to become larger and larger \cite{Domokos:2002,Asboth:2005}. This calls for developing a unifying theoretical formalism which allows one to describe the coupled atom-field dynamics below, at, and above the selforganization threshold. 

In this work we derive the Fokker-Planck equation governing the atomic dynamics, which is valid for any value of the intracavity field amplitude. This is obtained by following the procedure developed in Refs. \cite{Javanainen:1980,Javanainen2:1980,Dalibard:1985}, which allows us to derive an effective Fokker-Planck equation for the atomic motion, in which the cavity field is treated quantum mechanically. This treatment leads to equations of motion which can be simulated by means of stochastic differential equations \cite{Horak:JPB,Gardiner:Noise,Gardiner:Stochastic}. The numerical simulations allow for a reliable description of the dynamics without further assumptions on the state of the intracavity field. As an example, we determine the momentum distribution when the loss rates are much larger than the pump rate, so that the intracavity field is essentially in the vacuum. We then compare the predictions of our model on the atomic distribution with the predictions extracted with the model in Refs. \cite{Horak:JPB,Griesser}. 
This article is organized as follows. In Sec. \ref{Sec:system} the theoretical model is introduced and the effective master equation, describing the coupled dynamics of cavity and atoms, is obtained after eliminating the excited state of the atoms in second order perturbation theory. The Fokker-Planck equation for the atomic dynamics is derived in Sec. \ref{Sec:semiclassical}, and the numerical simulations of the center-of-mass motion dynamics are reported and discussed in Sec. \ref{Sec:numerical}. The conclusions are drawn in Sec. \ref{Sec:Conclusions}. The appendices report details of the calculations in Sec. \ref{Sec:semiclassical} and Sec. \ref{Sec:numerical}. 

\section{Theoretical model}
\label{Sec:system}

In this section we derive the master equation which is at the basis of the semiclassical treatment in Sec. \ref{Sec:semiclassical}. This effective master equation is obtained for a system of atoms inside the cavity and illuminated by a transverse pump, in the limit in which the atomic transition is driven below saturation. It thus describes the coupled dynamics of atomic center-of-mass motion and cavity field. 

\subsection{The system}

The physical system is illustrated in Fig. \ref{Fig:sys}: $N$ atoms are confined inside a resonator and their dipolar transitions scatter photons of a transverse laser and of the mode of a standing-wave cavity. The atoms are sufficiently hot to be considered distinguishable. For sake of generality we also assume that they can be of different species. We denote by $m_j$ the mass of atom $j=1,\ldots,N$, and by  $\ket{g}_j$ and $\ket{e}_j$ the ground and excited state of the corresponding dipolar transition, whose frequency $\omega_j$ is quasi-resonantly coupled with the laser and with the cavity field (the position dependence of the transition frequency also takes into account possible spatial inhomogeneities). The laser is here assumed to be a classical standing-wave field with frequency $\omega_L$ and wave vector perpendicular to the cavity axis, while the cavity mode is a quantum field of frequency $\omega_c$ and wave vector $k$. Cavity and laser field have the same linear polarization, so that they both drive the atomic dipolar transitions. We denote by $\hat {a}$ and $\hat{a}^{\dagger}$ the annihilation and creation operators of a cavity photon, with $[\hat{a},\hat{a}^{\dagger}]=1$. 

The center-of-mass motion of the atoms is restricted to the cavity axis, which here corresponds with the $x$-axis. The position and canonically conjugated momentum of atom $j$ are given by the operators $\hat{x}_j$ and $\hat{p}_j$, such that $[\hat{x}_j,\hat{p}_k]={\rm i}\hbar\delta_{jk}$ with $\delta_{jk}$ Kronecker delta. Internal and external degrees of freedom of the atoms couple via the mechanical effects of atom-photon interactions. Our purpose is to provide a theoretical description of the scattering dynamics leading to cooling of the atomic motion. 

We start with the master equation for the density matrix $\OPrho$ for the cavity and for the atoms internal and external degrees of freedom, which reads
\begin{eqnarray}
\label{masterrot}
\frac{\partial}{\partial t} \OPrho&=& - \frac{{\rm i}}{\hbar} \commutator{\OPham}{\OPrho} +  \mathcal{L}_{\kappa} \OPrho +  \mathcal{L}_{\gamma} \OPrho\\ 
&\equiv&\mathcal L\rho,
\end{eqnarray}
where $\mathcal L$ is the corresponding Lindbladian. Master equation \eqref{masterrot} is reported in the reference frame rotating at the laser frequency $\omega_L$. Here, the coherent dynamics are governed by Hamiltonian
\begin{align}
\OPham = &\sum_{j=1}^N \frac{\OPp_j ^2}{2 m_j} -\hbar \Delta_c \OPadagger \OPa - \sum_{j=1}^{N} \hbar \Delta_j \OPsigp_j \OPsigm_j \\ &+ \sum_{j=1}^N \hbar g_j(\OPx_j) (\OPadagger \OPsigm_j +  \OPsigp_j \OPa ) \nonumber  + \sum_{j=1}^N \hbar \Omega_j ( \OPsigp_j + \OPsigm_j)\,,
\end{align} 
where  $\OPsigm_j=\ket{g}_j\bra{e}$ and $\OPsigp_j$ is its adjoint, $\Delta_c = \omega_L - \omega_c$ and $\Delta_j = \omega_L - \omega_j$ are the detunings of the laser frequency from the cavity frequency $\omega_c$ and from the atomic transition frequency $\omega_j$, respectively; $\Omega_j$ is the real-valued coupling strength of atom $j$ with the laser, and $g_j(\OPx_j) = g_j \cos(k \OPx_j)$ is the real-valued coupling of the atomic transition, with $g_j$ the vacuum Rabi frequency for the atom $j$ and $\cos(k x)$ the spatial mode function.

The incoherent dynamics are assumed to be due to cavity losses, at rate $\kappa$, and to radiative decay of the atoms excited states, at rate $\gamma_j$. They are described by the superoperators 
\begin{align}
\mathcal{L}_{\kappa} \OPrho  &= - \kappa \big(\OPadagger \OPa \OPrho  + \OPrho  \OPadagger \OPa - 2 \OPa \OPrho  \OPadagger \big), 
\label{Lkappa}\\ 
\mathcal{L}_{\gamma} \OPrho  &= - \sum_{j=1}^N \frac{\gamma_j}{2} \big( \OPsigp_j \OPsigm_j \OPrho  + \OPrho  \OPsigp_j \OPsigm_j \big)+\mathcal{J}\hat{\rho}\,,
\label{Lgamma}
\end{align}
where
\beq
\mathcal{J}\hat{\rho} = \sum_{j=1}^N \gamma_j \int_{-1}^{1} {\rm d}u N_j(u) \ket{g}\bra{e}_j {\rm e}^{- {\rm i} k_j u \hat{x}_j}\hat{\rho} {\rm e}^{ {\rm i} k_j u \hat{x}_j} \ket{e}\bra{g}_j\,.
\eeq
This term describes the jump from $\ket{e}_j$ to $\ket{g}_j$ due to spontaneous decay and takes into account the momentum transfer along the cavity axis to the atom due to spontaneous emission \cite{Cohen-Tannoudji}. Here, the dipole radiation pattern $N_j(u)$ is normalized and symmetric about $u=0$. For later convenience, we define its second moment by $(\overline{u^2})_j$, such that
\begin{align}
\int_{-1}^{1} {\rm d}u N_j(u) u^2 = (\overline{u^2})_j\,.
\end{align} 

\subsection{Adiabatic elimination of the excited state}
\label{adiabatic}

We now proceed in deriving the master equation for cavity field and atoms center-of-mass motion when the occupation of the atomic excited states can be neglected. Let us first assume that the particles do not move. In this case their coupling $g_j$ with the cavity field is fixed, and the excited state can be eliminated in second order in an expansion in the parameter $1/|\Delta_j|$, assuming that $|\Delta_j| \gg \gamma_j/2,\Omega_j$ and $|\Delta_j| \gg |\Delta_c|,\kappa,g_j \sqrt{\bar{n}}$, with  $\bar{n}$ is the mean photon number in the cavity. For $N$ atoms, the condition on the coupling strengths becomes $|\Delta_j| \gg \sqrt{N} \Omega_j,\sqrt{N} (g_j \sqrt{\bar{n}})$, see \cite{Habibian:2011}. When the center-of-mass motion is considered, on the other hand, the coupling strength $g_j$ varies as a function of time. Moreover, atoms with different velocities experience different Doppler shifts, which modify the resonance condition. These effects can be neglected when the corresponding time scales are longer than the typical time scale in which the excited state is occupied, i.e., when $k\bar{p_j}/m_j\ll |\Delta_j|$ (with $\bar{p_j}=\sqrt{\mean{p_j^2}}$), which is satisfied when the atomic gas has previously been Doppler cooled \cite{Stenholm:1986,Castin:1989}.  

Formally, the effective master equations describing the dynamics of cavity field and atoms center of mass motion is obtained by deriving a closed equation of motion for the reduced density operator $\OP{v}$ when the atoms are all in the internal ground state $\ket{\bm{g}}= \ket{g_1, g_2, ..., g_N}$. The reduced density operator $\hat{v}$ is defined as $\hat{v}=P\hat{\rho}$, where 
\begin{align}
\OP{v} &= P \OP{\rho} = \ket{\bm{g}}\bra{\bm{g}} \bra{\bm{g}} \OP{\rho} \ket{\bm{g}}\,, 
\label{opv} 
\end{align}
such that $\hat{\rho}=\hat{v}+\hat{w}$ with
\begin{align}
\OP{w} &= Q \OP{\rho}\,,
\label{opw}
\end{align}
where $P$ and $Q=1-P$ are projectors ($P^2=P$, $Q^2=Q$, $P^\dagger=P$, $Q^{\dagger}=Q$) and $(P + Q)\hat{\rho} = \hat{\rho}$. In order to adiabatically eliminate the excited state, we rewrite the Lindbladian as $\mathcal L=(P+Q)\mathcal L(P+Q)$ and introduce the decomposition
\begin{align}
\mathcal L= \mathcal{L}^A + \mathcal{L}^F + \mathcal{J} + \mathcal{L}^{\text{int}}\,,
\end{align}
where 
\beq
\mathcal{L}^F\hat{\rho}=-\frac{\rm i}{\hbar}\left[ \sum_{j=1}^N\frac{\hat{p}_j^2}{2m_j}- \hbar \Delta_c\hat{a}^\dagger\hat{a},\OPrho\right] +\mathcal L_{\kappa}\hat{\rho}\,
\eeq
with $\mathcal{L}^FP=P\mathcal{L}^F$. Term 
\beq
\mathcal{L}^A \OPrho = \sum_{j=1}^N \left( {\rm i}\Delta_j\left[ \ket{e}_j\bra{e},\OPrho\right] - \frac{\gamma_j}{2} (\ket{e}\bra{e}_j \OPrho + \OPrho \ket{e}_j\bra{e} )\right)\,
\eeq
is such that $Q\mathcal{L}^AQ=\mathcal{L}^A$ and $\mathcal{L}^AP=P\mathcal{L}^A=0$, while $\mathcal{J}P=0$ and $P\mathcal{J}=P\mathcal{J}Q$. Finally, $P\mathcal{L}^{\text{int}}P=0$, with 
\begin{eqnarray}
\mathcal{L}^{\text{int}} \hat{\rho} = -{\rm i}\sum_{j=1}^N \commutator{\left(\ketbra{e}{g}_j(\Omega_j+
g_j(\OPx_j) \hat{a})+{\rm H.c}\right)}{\hat{\rho}}\,.
\end{eqnarray}
The master equation \eqref{masterrot} is thus rewritten in terms of coupled differential equations for the time evolution of $\OP{v}$ and $\OP{w}$ defined in Eqs. \eqref{opv}-\eqref{opw}. The formal solution for $\OP{w}$ reads 
\begin{align}
\hat{w}(\tau) &=  
                                  \int_{0}^{\tau} {\rm d}\tau' {\rm e}^{Q (\mathcal{L}^A + \mathcal{L}^F) (\tau - \tau')} Q \mathcal{L}^{\text{int}} \hat{w}(\tau') 
                                  \nonumber \\ &+ \int_{0}^{\tau} {\rm d}\tau' {\rm e}^{Q (\mathcal{L}^A + \mathcal{L}^F) (\tau - \tau')} Q \mathcal{L}^{\text{int}} \hat{v}(\tau')\,,
\label{wtau}                                  
\end{align}
where we have assumed $\hat{w}(0)=0$, namely, all atoms are in the internal ground state at $t=0$. Using Eq. \eqref{wtau} in the differential equation for $\OP{v}$ leads to an integrodifferential equation of motion
\begin{align}
\label{pert}                              
\frac{\partial}{\partial t} \OP{v} &= P \mathcal{L}^F \OP{v}+ P (\mathcal{J} + \mathcal{L}^{\text{int}}) \\
                                   &\times\int_{0}^{t} {\rm d}\tau {\rm e}^{Q (\mathcal{L}^A + \mathcal{L}^F) (t - \tau)} Q \mathcal{L}^{\text{int}} [\OP{v}(\tau)+\OP{V}(\tau)]  \,,\nonumber
\end{align}  
with
\beq                              
\OP{V}(\tau) = \int_{0}^{\tau} {\rm d}\tau' {\rm e}^{Q (\mathcal{L}^A + \mathcal{L}^F) (\tau - \tau')} Q \mathcal{L}^{\text{int}}\left[ \OP{v}(\tau')+\OP{w}(\tau')\right]\,.\nonumber
\eeq
Equation \eqref{pert} can be brought to the form of an effective Born-Markov master equation by performing a perturbative expansion to the second order in the small parameters $\epsilon_{\text{int}} \propto \sqrt{N}\frac{g \sqrt{\bar{n}}}{|\Delta_a|}, \sqrt{N}\frac{\Omega}{|\Delta_a|}$ and $\epsilon_{F} \propto \frac{|\Delta_c|}{|\Delta_a|}, \frac{\kappa}{|\Delta_a|}$, which consists in neglecting terms like $(\mathcal{L}^{\text{int}})^3$, $(\mathcal{L}^{\text{int}})^2 \mathcal{L}^{F}$, $\mathcal{L}^{\text{int}} (\mathcal{L}^{F})^2 $, $ (\mathcal{L}^{F})^3$, and higher. In this approximation the master equation for the reduced density matrix $\OP{v}$, Eq. \eqref{pert} is reduced to the form
\begin{eqnarray}                                     
\label{masteradiab}                                     
&&\frac{\partial}{\partial t} \OP{v} = -\frac{\rm i}{\hbar} \commutator{\OPham_{\text{eff}}}{\OP{v}} - \kappa (\OPadagger \OPa \OP{v} + \OP{v} \OPadagger \OPa - 2 \OPa \OP{v} \OPadagger) \\
&&-\sum_{j=1}^N \frac{\gamma_j^\prime}{2} \Big\{ \hat{B}_j^\dagger\hat{B}_j\OP{v}+\OP{v}\hat{B}_j^\dagger\hat{B}_j -2 \int_{-1}^{1} {\rm d}u N_j(u)  \hat{B}_j \OP{v} \hat{B}_j^\dagger\Big\}\,,
\nonumber
\end{eqnarray}
with $\gamma_j^\prime=\gamma_jg_j^2/(\Delta_j^2 + \gamma_j^2/4)$ the rate of incoherent photon scattering via spontaneous decay, while operator
\begin{equation}
\hat{B}_j={\rm e}^{- {\rm i} k_j u \OPx_j}\left(\cos (k \OPx_j) \OPa + \frac{\Omega_j}{g_j}\right)\,
\end{equation}
describes the mechanical effect associated with absorption of a laser or a cavity photon and followed by a spontaneous emission. The effective Hamiltonian $\OPham_{\text{eff}}$ reads
\begin{align}
\label{Hameff}
\OPham_{\text{eff}} = &\sum_{j=1}^N \frac{\OPp_j ^2}{2 m_j} - \hbar \left( \Delta_c - \sum_{j=1}^N U_j \cos^2(k \OPx_j) \right) \OPadagger \OPa \nonumber \\ &+ \hbar\sum_{j=1}^N S_j\cos(k \OPx_j) (\OPa + \OPadagger)\,.
\end{align}
It contains the shift of the cavity frequency due to the interaction with the atoms, which scales with the frequency
\beq
U_j= \frac{\Delta_j g_j^2}{\Delta_j^2 + \gamma_j^2/4}\,,
\eeq
and the pump on the cavity field due to coherent scattering into the cavity mode, which scales with the amplitude
\beq
S_j= \Delta_j \frac{g_j \Omega_j}{\Delta_j^2 + \gamma_j^2/4} \,.
\eeq
The corresponding terms in Eq. \eqref{Hameff} depend on the atomic positions and give rise to mechanical forces on the atoms.  We remark that the master equation in Eq. \eqref{masteradiab} has been reported, for instance, in Refs. \cite{Horak:JPB,Asboth:2005}, where the internal dynamics are eliminated by setting $\hat{\sigma}_z \approx -1$ and ultimately expressing $\OPsigm_j$ and $\OPsigp_j$ in terms of cavity field operators. Here, we have given its detailed derivation using second-order perturbation theory by means of projectors acting on density operators \cite{Cirac:1992}.

\section{Semiclassical model}
\label{Sec:semiclassical}

In this section we analyse the predictions of Eq. \eqref{masteradiab} under the assumption that the atoms center-of-mass motion can be treated semiclassically. For this purpose we first consider the dynamics of the atoms in Wigner representation and denote by $\OP{W} _t(\bm{x}, \bm{p})$ the operator for the cavity field degrees of freedom, where the subscript $t$ indicates the time. Operator $\OP{W} _t(\bm{x}, \bm{p})$ is related to the reduced density operator $\hat{v}$ by the equation
\beq
\label{Wigner:1}
\hat{W}_t (\bm{x}, \bm{p}) = \frac{1}{(2 \pi \hbar)^N} \int_{-\infty}^{\infty} {\rm d}\bm{\xi} {\rm e}^{- \frac{i}{\hbar} \bm{p} \cdot \bm{\xi}} \bra{\bm{x} + \frac{1}{2} \bm{\xi}} \hat{v} \ket{\bm{x} - \frac{1}{2} \bm{\xi}}\,,
\eeq
with  $\bm{y}= (y_1, \ldots, y_N)$, where $\bm{y}=\bm{x},\bm{p},\bm{\xi}$. It is a scalar function of the atoms positions $x_j$ and canonically conjugated momenta $p_j$ of the atoms. Note that $\bm{p} \cdot \bm{\xi} = \sum_{j=1}^N p_j \xi_j $ and $\int_{-\infty}^{\infty} {\rm d}\bm{\xi} = \prod_{j=1}^N \int_{-\infty}^{\infty} {\rm d}\xi_j$. The Wigner function for the atoms is denoted by $f(\bm{x}, \bm{p}, t)$ and is defined as
\begin{align}
\label{wigner:f}
f(\bm{x},\bm{p},t) = \text{Tr} \left\{ \OP{W}_t (\bm{x}, \bm{p}) \right\}\,.
\end{align}
From the density operator in Eq. \eqref{Wigner:1} one can find the  combined atom-field  Wigner function $W (\bm{x}, \bm{p}, \alpha, \alpha^{*})$ used in Ref. \cite{Horak:JPB} by means of the relation
\begin{align}
W _t(\bm{x}, \bm{p}, \alpha, \alpha^{*}) =   \int \frac{{\rm d}^2 \eta}{\pi ^2} {\rm e}^{\eta ^{*} \alpha - \eta \alpha ^{*}} \text{Tr}\left\{ \hat{W}_t(\bm{x}, \bm{p}) \hat{\mathcal{D}}(\eta) \right\} \,,
\label{combwig}
\end{align}
where $\alpha$ and $\alpha^*$ are the variables for the cavity field amplitude and $\hat{\mathcal{D}}(\eta)=\exp\left(\eta\hat{a}^\dagger-\eta^*\hat{a}\right)$ 
is the displacement operator for the cavity field, with $\eta$ complex variables. 

Let us now discuss the conditions under which the motion can be treated as a semiclassical variable. This is possible when the typical width of the momentum distribution, which we denote by $\Delta p_j$ for the atom $j$, is much larger than the photon momentum $\hbar k$, 
\begin{align}
\hbar k \ll \Delta p_j\,.
\label{cond1}
\end{align}
In this limit the momentum changes due to emission and absorption of a photon are very small. In addition, the uncertainty in the atomic position, $\Delta x_j$, is larger than the value set by the Heisenberg uncertainty relation, $\Delta x_j>\hbar/\Delta p_j$. These conditions are met when the atoms are at the stationary state of Doppler cooling, such that  $\Delta p_j^2/(2m_j)\sim \hbar \gamma_j/4$ when $\gamma_j\gg \omega_r$, with $\omega_r = \frac{\hbar k^{2}}{2 m}$ the recoil frequency \cite{Stenholm:1986,Castin:1989}. In this work we derive a Fokker-Planck equation starting from this assumption, and then check that the corresponding stationary state fulfills the conditions under which the Fokker-Planck equation is valid.

\subsection{Semiclassical model below the selforganization threshold}

We now derive equations of motion for the atomic degrees of freedom by eliminating the cavity degrees of freedom. This is possible provided the cavity degrees of freedom evolve on a faster time scale than the atomic motion, namely, when 
\beq
\label{cond:2}
k\Delta p_j/m_j \ll |\kappa+{\rm i}\Delta_c |\,.
\eeq
As one can easily check, this condition is consistent with Eq. \eqref{cond1}, provided that $\omega_r \ll \kappa$. The following treatment extends the method applied in Ref. \cite{Dalibard:1985} to the dynamics of atoms coupled to a resonator. We start with the master equation in Eq. \eqref{masteradiab} in Wigner representation for the atomic degrees of freedom and consider the reference frame moving with the atoms and defined by the relation
\begin{align}
\hat{\tilde{W}}_t (\bm{x},\bm{p}) = \hat{W}_t (\bm{x} +\bm{v} (t-t_0) ,\bm{p})\,,
\label{movingframe}
\end{align}
where ${\bm v}=(p_1/m_1,\ldots,p_N/m_N)$, $\hat{W} _t(\bm{x},\bm{p}) $ is defined in Eq. \eqref{Wigner:1} and $\hat{\tilde{W}}_t (\bm{x},\bm{p}) $ is given in the reference frame moving with the atoms, with $t_0$ an initial time. Its time evolution reads 
\begin{widetext}
\begin{eqnarray}
\label{Wigner:Below} 
\frac{\partial}{\partial t} \hat{\tilde{W}}_t \begin{pmatrix} \bm{x} \\ \bm{p}  \end{pmatrix} &=& 
\mathcal{L}_0' \hat{\tilde{W}}_t \begin{pmatrix} \bm{x} \\ \bm{p}  \end{pmatrix} 
+ \mathcal{L}^{\gamma} \hat{\tilde{W}}_t \begin{pmatrix} \bm{x} \\ \bm{p}  \end{pmatrix}
\\
&-&\frac{\rm i}{2} \sum_j S_j
\Bigl[(\OPa + \OPadagger) \bigg ({\rm e}^{ {\rm i} k (x_j + \frac{p_j}{m_j}\tau)} \hat{\tilde{W}}_t \begin{pmatrix} \bm{x} +\tau\hbar \bm{k_j}/(2m_j)  \\ \bm{p} - \hbar \bm{k_j}/2   \end{pmatrix} 
+ {\rm e}^{- {\rm i} k (x_j + \frac{p_j}{m_j} \tau) } \hat{\tilde{W}}_t \begin{pmatrix} \bm{x} -\tau\hbar \bm{k_j}/(2m_j)  \\ \bm{p} + \hbar \bm{k_j}/2   \end{pmatrix} \bigg) \nonumber\\
&& \qquad \quad - \bigg( {\rm e}^{ {\rm i} k (x_j + \frac{p_j}{m_j} \tau) } \hat{\tilde{W}}_t \begin{pmatrix} \bm{x} -\tau\hbar \bm{k_j}/(2m_j)  \\ \bm{p}+ \hbar \bm{k_j}/2   \end{pmatrix} 
+ {\rm e}^{- {\rm i} k (x_j + \frac{p_j}{m_j} \tau) } \hat{\tilde{W}}_t \begin{pmatrix} \bm{x} + \tau\hbar \bm{k_j}/(2m_j)   \\ \bm{p}- \hbar \bm{k_j}/2  \end{pmatrix} \bigg) (\OPa + \OPadagger)\Bigr] \nonumber 
\\
&-&\frac{\rm i}{4} \sum_j U_j  
\Bigl[\OPadagger \OPa \bigg( {\rm e}^{2 {\rm i} k (x_j + \frac{p_j}{m_j}\tau)} \hat{\tilde{W}}_t \begin{pmatrix} \bm{x} + \tau\hbar \bm{k_j}/m_j  \\ \bm{p} - \hbar \bm{k_j}   \end{pmatrix} + {\rm e}^{-2 {\rm i} k (x_j + \frac{p_j}{m_j} \tau) } \hat{\tilde{W}}_t \begin{pmatrix} \bm{x} -  \tau\hbar \bm{k_j}/m_j \\ \bm{p} + \hbar \bm{k_j}  \end{pmatrix} + 2 \hat{\tilde{W}}_t \begin{pmatrix} \bm{x} \\ \bm{p} \end{pmatrix} \bigg) 
\nonumber \\
&& \qquad \quad - \bigg( {\rm e}^{2 {\rm i} k (x_j + \frac{p_j}{m_j} \tau) } \hat{\tilde{W}}_t \begin{pmatrix} \bm{x} -  \tau\hbar \bm{k_j}/m_j \\ \bm{p}+\hbar \bm{k_j} \end{pmatrix}+ {\rm e}^{-2 {\rm i} k (x_j + \frac{p_j}{m_j} \tau) } \hat{\tilde{W}}_t \begin{pmatrix} \bm{x} + \tau\hbar \bm{k_j}/m_j \\ \bm{p}- \hbar \bm{k_j} \end{pmatrix}+ 2 \hat{\tilde{W}}_t \begin{pmatrix} \bm{x} \\ \bm{p} \end{pmatrix}  \bigg) \OPadagger \OPa \Bigr]\,,\nonumber 
\end{eqnarray}
where $\hat{\tilde{W}}_t (\bm{x},\bm{p}) \equiv \hat{\tilde{W}}_t \begin{pmatrix} \bm{x} \\ \bm{p}  \end{pmatrix}$, $\tau = t- t_0$, and $(\bm{k_j})_\ell=k\delta_{\ell,j}$, while
\begin{eqnarray}
\mathcal{L}_0' \hat{\tilde{W}}_t \begin{pmatrix} \bm{x} \\ \bm{p}  \end{pmatrix} = {\rm i} \Delta_c \left[\hat{a}^{\dagger}\hat{a},\hat{\tilde{W}}_t \begin{pmatrix} \bm{x} \\ \bm{p}  \end{pmatrix}\right]+ \kappa \left(2\hat{a} \hat{\tilde{W}}_t \begin{pmatrix} \bm{x} \\ \bm{p}  \end{pmatrix} \hat{a}^\dagger -  \hat{a}^{\dagger}\hat{a}\hat{\tilde{W}}_t \begin{pmatrix} \bm{x} \\ \bm{p}  \end{pmatrix} -  \hat{\tilde{W}}_t \begin{pmatrix} \bm{x} \\ \bm{p}  \end{pmatrix}\hat{a}^{\dagger}\hat{a}\right)\,.
\end{eqnarray}
The effects of spontaneous emission are included in $\mathcal{L}^{\gamma} \hat{\tilde{W}}_t \begin{pmatrix} \bm{x} \\ \bm{p}  \end{pmatrix}$ , whose detailed form is reported in Appendix A. 
\end{widetext}
After performing a Taylor expansion till second order in the parameters $\epsilon_1=\hbar k/\Delta p$ and $\epsilon_2 = \frac{k \Delta p}{m \kappa}$, Eq. \eqref{Wigner:Below} can be cast in the form
\begin{align}
\frac{\partial}{\partial t} \hat{\tilde{W}}_t ( \bm{x}, \bm{p}) = \left(\mathcal{L}_0+ \mathcal{L}_1 (t) + \mathcal{L}_2 (t) \right)
\hat{\tilde{W}}_t ( \bm{x}, \bm{p})\,,\label{tildeW-exp}
\end{align}
where $\mathcal{L}_j$ is of $j$-th order in $\epsilon_1,\epsilon_2$ and we have assumed $\epsilon_1 \sim \epsilon_2 \sim \epsilon$, which is correct provided that $\omega_r \ll \kappa$. Operators $\hat{\tilde{W}}_t ( \bm{x}, \bm{p})$ appear now all at the same positions and momenta $\bm{x}, \bm{p}$, so that we omit to write the argument explicitly. Superoperators $\mathcal{L}_j$ are defined as 
\begin{eqnarray}
\label{L0:gamma}
\mathcal{L}_0 \hat{\tilde{W}}_t   
&=& 
\mathcal{L}_0' \hat{\tilde{W_t}}  - {\rm i}  \sum_j U_j \left[\OPadagger\OPa,\cos ^2(kx_j) \hat{\tilde{W}}_t \right]\\
& &- {\rm i}  \sum_j S_j \left[ (\OPa + \OPadagger),\cos (kx_j) \hat{\tilde{W}}_t  \right] + \mathcal{L}_0^{\gamma'} \hat{\tilde{W}}_t \,,\nonumber \\
\label{L1:gamma}
\mathcal{L}_1\hat{\tilde{W}}_t  &=&
+\frac{\rm i}{\hbar} \tau \sum_j \left[ \frac{p_j}{m_j} \hat{F}_j , \hat{\tilde{W}}_t  \right]  \\
& &- \frac{1}{2} \sum_j \Big\{ 
( \frac{\partial}{\partial p_j} - \frac{\tau}{m_j} \frac{\partial}{\partial x_j})
\hat{\tilde{W}}_t , \hat{F}_j    \Big\}_+ + \mathcal{L}_1^{\gamma'}\hat{\tilde{W}}_t  \,,\nonumber \\
\label{L2:gamma}
\mathcal{L}_2\hat{\tilde{W}} &=&   \frac{ {\rm i} \hbar}{8} \sum_j\left[\frac{\partial^2}{\partial p_j^2}\hat{\tilde{W}}_t,\frac{\partial}{\partial x_j} \hat{F}_j  \right] + \mathcal{L}_2^{\gamma'} \hat{\tilde{W}}_t \,,
\end{eqnarray}
with $\hat{F}_j$ the force operator on the $j$-th atom \cite{Cirac:1992}, which is defined as
\begin{align*}
\hat{F}_j = &\hbar k S_j \sin(kx_j) (\hat{a} + \hat{a}^\dagger) +\hbar k U_j \sin(2 k x_j) \hat{a}^\dagger \hat{a} \\
                 &-{\rm i}  (\hat{a}^\dagger - \hat{a})\hbar k \frac{\gamma_j'}{2} s_j \sin(kx_j)\,
\end{align*}
with $s_j = \Omega_j / g_j$. The terms $\mathcal{L}_0^{\gamma'} \hat{\tilde{W}}, \mathcal{L}_1^{\gamma'} \hat{\tilde{W}}$ and $\mathcal{L}_2^{\gamma'} \hat{\tilde{W}}$ are due to spontaneous emission and their explicit form is given in Appendix A. Note that $\mathcal{L}_2$ is evaluated at $\tau=0$ since this term is already of second order in $\epsilon$.\\

We rewrite the operator $\tilde{\hat{W}}_t$ as
\begin{align}
\label{eq:ad}
\tilde{W}_t (\bm{x}, \bm{p}) = \tilde{f} (\bm{x}, \bm{p}, t) \sigma_s(\bm{x}) + \tilde{\chi} (\bm{x}, \bm{p}, t)
\end{align}
where $\sigma_s(\bm{x})$ is the density matrix for the field, which solves equation $ \mathcal{L}_0\sigma_s(\bm{x})=0$ for $N$ atoms fixed at positions $x_j$ while $\tilde{f} (\bm{x}, \bm{p}, t)$ is the Wigner function of Eq.  \eqref{wigner:f} in the reference frame moving with the atom. Therefore, $\tilde{f} (\bm{x}, \bm{p}, t) \sigma_s(\bm{x})$ corresponds to the solution, in which the cavity field follows adiabatically the external atomic motion, while the non-adiabatic terms are contained in the (traceless) operator $\tilde{\chi}=\tilde{W}-{\rm Tr}\{\tilde{W}\}\sigma_s$. When condition \eqref{cond:2} is fulfilled, this contribution is expected to be a small correction and reads
\begin{eqnarray}
\label{eq:chi}
\tilde{\chi}(t)&=&\int_{t_0}^t{\rm d}t'{\rm e}^{{\mathcal L}_0(t-t')}({\mathcal L}_1(t')f(\bm{x},\bm{p},t')\sigma_s\\
&&-{\rm Tr}\{{\mathcal L}_1(t')f(\bm{x},\bm{p},t')\sigma_s\}\sigma_s)\,,\nonumber
\end{eqnarray}
where the value at $t=t_0$ is taken to be zero, as is consistent with the assumption that when the transverse laser is switched on there are no correlations between field and atoms. Under this assumption, we use Eq. \eqref{eq:ad} in Eq. \eqref{tildeW-exp} and consider a coarse-grained dynamics. Applying the Markov approximation after tracing over the cavity degrees of freedom, we obtain the equation governing the time evolution of the Wigner function $\tilde{f}(\bm{x},\bm{p},t)$, which is valid till second order in $\epsilon$:
\begin{widetext}
\begin{align}
\label{timef}\\
\frac{\partial}{\partial t} &\tilde{f}  (\bm{x},\bm{p},t) \Big|_{t=t_0} =\text{Tr} \left\{\Big( \mathcal{L}_1 (\bm{x}, \bm{p},  t_0) + \mathcal{L}_2 (\bm{x}, \bm{p}, t_0) \Big)  \cdot\tilde{f} (\bm{x}, \bm{p},  t_0) \sigma_s(\bm{x}) \right\}\nonumber \\
+ &\text{Tr} \left\{ \mathcal{L}_1 (\bm{x}, \bm{p},  t_0)
\int_{- \infty}^{t_0} {\rm d}t' {\rm e}^{\mathcal{L}_0 ( t_0-t')} 
\Big[ \mathcal{L}_1 (\bm{x}, \bm{p}, t') \cdot \tilde{f}(\bm{x}, \bm{p},  t_0) \sigma_s(\bm{x}) - \text{Tr} \left\{ \mathcal{L}_1 (\bm{x}, \bm{p}, t') \cdot \tilde{f}(\bm{x}, \bm{p},  t_0) \sigma_s(\bm{x}) 
\right\}\sigma_s(\bm{x}) \Big] \right\} \nonumber .
\end{align}
\end{widetext}
The equation in the original reference frame is found by using the relation $\tilde{f}(\bm{x},\bm{p},t_0) = f(\bm{x},\bm{p},t_0) $ together with equation
\begin{align*}
\frac{\partial}{\partial t} \tilde{f} |_{t=t_0} = \frac{\partial}{\partial t} f |_{t=t_0} + \bm{v} \cdot \frac{\partial}{\partial \bm{x}} f |_{t=t_0}\,.
\end{align*} 
After observing that the trace over the term containing operator $\mathcal{L}_2$ (neglecting $\mathcal{L}_2^{\gamma'}$ ) vanishes, we cast Eq. \eqref{timef} in the form 
\begin{widetext}
\begin{align}
\label{FPEsemi}
\frac{\partial}{\partial t} f(\bm{x},\bm{p},t) |_t  
=& - \sum_{j=1}^N  \frac{\partial}{\partial x_j} \frac{p_j}{m_j} f(\bm{x},\bm{p},t)  
- \sum_{j=1}^N \frac{\partial}{\partial p_j} \left( \Phi_j -\sum_{\ell=1}^N \gamma_{j\ell}  \, p_\ell\right) f (\bm{x}, \bm{p},t)   \\
&+\sum_{j,\ell=1}^N \frac{\partial^2}{\partial p_j \partial p_\ell}  D_{j\ell}   f(\bm{x}, \bm{p},  t) + \sum_{j,\ell=1}^N \frac{\partial}{\partial p_j}  \eta_{j\ell}  \frac{\partial}{\partial x_\ell} f(\bm{x}, \bm{p},  t)\,, \nonumber
\end{align}
where the derivatives are now explicitly reported. This equation has the form of a Fokker-Planck equation for the atomic center-of-mass variables, while the field enters in the coefficients through the expectation values of field variables taken over the density matrix $\sigma_s(\bm{x})$. In particular, $ \Phi_j  = \text{Tr} \left\{ \sigma_s (\bm{x}) \OP{F}_j \right\}$ is the mean dipole force over the $j$ atom due to the cavity field, and $\gamma_{j\ell}$  are the friction coefficients which read
\begin{align*}
\gamma_{j\ell}  =   \gamma_{j\ell}'+\text{Tr} \left\{ \OP{F}_j \int_{ 0}^{\infty} {\rm d}\tau  \exp\left(\mathcal{L}_0 \tau\right) \frac{ {\rm i} \tau}{ \hbar m_\ell} [ \OP{F}_\ell ,\sigma_s(\bm{x})] \right\}\,,    
\end{align*}
where $\gamma_{j\ell}'$ is the contribution due to spontaneous emission while the second term arises from the coupling with the cavity. Coefficients $D_{j\ell}$ are the diffusion matrix coefficients, they include the contribution due to spontaneous decay ($D_{j\ell}'$) and the contribution to diffusion due to the cavity field, 
\begin{align*}
 D_{j\ell}  = D_{j\ell}'+\text{Tr} \left\{ \OP{F}_j \int_{ 0}^{\infty} {\rm d}\tau \exp\left(\mathcal{L}_0 \tau\right)\Big( \frac{1}{2} \big\{ \sigma_s(\bm{x} ), \OP{F}_\ell \big\}_+ - \text{Tr} \left\{ \sigma_s (\bm{x}) \OP{F}_\ell \right\} \sigma_s(\bm{x}) \Big) \right\}\,. 
\end{align*}
Finally, the Fokker-Planck equation exhibits cross derivatives between position and momentum of the particles with coefficients
\begin{align*}
 \eta_{j\ell}  = \eta_{j\ell}' +\text{Tr} \left\{ \OP{F}_j \int_{ 0}^{\infty} {\rm d}\tau  \exp\left(\mathcal{L}_0 \tau\right)\frac{\tau}{m_{\ell} } \Big( \frac{1}{2} \big\{ \sigma_s(\bm{x} ), \OP{F}_\ell \big\}_+ - \text{Tr} \left\{ \sigma_s (\bm{x}) \OP{F}_\ell \right\} \sigma_s(\bm{x}) \Big) \right\}\,, 
\end{align*}
\end{widetext}
where the first term, $\eta_{j\ell}'$, is due to spontaneous emission. This term can be rewritten as
\begin{align*}
\sum_{j,\ell=1}^N \frac{\partial}{\partial p_j}  \eta_{j\ell}  \frac{\partial}{\partial x_\ell} f 
=  \sum_{j,\ell=1}^N \frac{\partial}{\partial p_j} \Big( 
 \frac{\partial}{\partial x_\ell} [\eta_{j\ell} f ]-  [ \frac{\partial \eta_{j\ell}}{\partial x_\ell} ] f
 \Big) ,
\end{align*}
where the second term in the parentheses gives a contribution to the force of higher order in $\epsilon$ and can thus be discarded \cite{Javanainen:1980,Javanainen2:1980}. 
The other term can also be neglected well below the selforganization threshold, when the spatial distribution has width which largely exceeds the cavity wave length $\lambda$ \cite{Dalibard:1985}. It must be taken into account, nevertheless, at and above the selforganization threshold, when spatial structures with periodicity $\lambda$ form. 

We remark that the explicit form of the coefficients due to spontaneous emission, $\gamma_{j\ell}', D_{j\ell}' , \eta_{j\ell}' $, is reported in Appendix A. These coefficients characterize the dynamics also in absence of the resonator. In the limit which we will consider here, where the laser and cavity fields are far detuned from the atomic resonance, they are of higher order and their contribution to the dynamics can often be discarded. 

We finally give the form of the field density matrix $\sigma_s(\bm{x})$. By solving $\mathcal{L}_0 \sigma_s(\bm{x})=0$ we find $\sigma_s(\bm{x}) = \ket{\alpha(\bm{x})}\bra{\alpha(\bm{x})}$, with $ \ket{\alpha(\bm{x})}$ coherent state of amplitude
\beq
\alpha(\bm{x}) = \frac{\sum_j S_j \big[ 1 - {\rm i} (\gamma_j/2\Delta_j ) \big] \cos(kx_j)}
{\big[ \Delta_c - \sum_j U_j \cos^2(kx_j) \big] +{\rm i} \kappa'}\,,
\label{alpha:x}
\eeq
with
\begin{align*}
\kappa' = \kappa + \sum_j \frac{g_j^2}{(\gamma_j/2)} \left( \frac{\gamma_j}{2\Delta_j}\right)^2 \frac{\Delta_j^2}{\Delta_j^2 + \gamma_j^2/4} \cos^2(kx_j) \,.
\end{align*}
Operators of the form $\mathcal F(a,a^{\dagger})$, which are function of the field variables, have expectation value
\beq
\label{n:cav}
\langle \mathcal F\rangle =\int {\rm d}\bm{x} {\rm d}\bm{p} {\rm Tr}\{W_t(\bm{x}, \bm{p})\mathcal F\}\,,
\eeq
where $W_t$ is found from Eq. \eqref{eq:ad} using the non-adiabatic term  in Eq. \eqref{eq:chi} after applying the Markov approximation.\\
\\

\subsection{Comparison with the semiclassical model in \cite{Horak:JPB}}

We now consider the Fokker-Planck equation derived in Ref. \cite{Horak:JPB}. This is based on the assumption that both atomic motion and cavity field can be treated semiclassically. With respect to the previous treatment, hence, here one also assumes that the mean field amplitude is large, $\langle \hat a\rangle=| \alpha_{0} | \gg 1$, so that one can perform an expansion in the quantum fluctuations about the mean value $ \alpha_{0}$. This allows one to discard higher derivatives in the field and atomic variables, thereby obtaining a Fokker-Planck equation. 

In this regime, it is convenient to consider the Wigner function for field and atomic motion in Eq. \eqref{combwig}, whose time evolution is given by master equation \eqref{masteradiab} in Wigner representation. The corresponding Fokker-Planck equation in the semiclassical limit is obtained by performing an expansion till second order in the small parameters $\epsilon_{1,j}= \frac{\hbar k}{(\Delta p)_j}$ and $\epsilon_2= \frac{1}{|\alpha_0 |}$, where it is assumed that $\epsilon_{1,j}$ and $\epsilon_{2}$ are approximately of the same order. The resulting time evolution reads
\begin{widetext}
\begin{align}
\label{FPEend}   
\frac{\partial}{\partial t} W_t
= &- \frac{\partial}{\partial \alpha_r} \bigg[ - \Delta_c' \alpha_i - \kappa' \alpha_r -  \sum_{j=1}^N s_j \Gamma_j  \cos(k x_j) \bigg] W_t- \frac{\partial}{\partial \alpha_i} \bigg[  \Delta_c' \alpha_r - \kappa' \alpha_i  - \sum_{j=1}^N s_j U_j \cos(k x_j) \bigg] W_t   \\
  &-  \sum_{j=1}^N \hbar\frac{\partial}{\partial p_j} \nabla_j \bigg[ - U_j |\alpha|^2  \cos^2(k x_j) - s_j U_j  \cos(k x_j) (2 \alpha_r) \nonumber+ (2 \alpha_i) s_j \Gamma_j  \cos(k x_j) \bigg] W_t   - \sum_{j=1}^N \frac{\partial}{\partial x_j} \bigg[ \frac{p_j}{m_j} \bigg] W_t  \nonumber \\
 & +  \frac{1}{4}\left(\frac{\partial^2}{\partial \alpha_r ^2} + \frac{\partial^2}{\partial \alpha_i ^2}\right)  \kappa'  W_t + \sum_{j=1}^N \frac{\hbar k}{2} \Gamma_j  \sin(2 k x_j) \frac{\partial}{\partial p_j}\left( \alpha_i \frac{\partial}{\partial \alpha_r} -\alpha_r \frac{\partial}{\partial \alpha_i}\right)W_t\nonumber \\ 
& + \sum_{j=1}^N (\hbar k)^2\Gamma_j \frac{\partial^2}{\partial p_j^2} \bigg[  |\alpha|^2  [ \sin^2(k x_j) +  (\overline{u^2})_j \cos^2( kx_j) ]  + s_j  (\overline{u^2})_j (2 \alpha_r \cos(k x_j) +s_j) \bigg] W_t \nonumber
\end{align}
\end{widetext}
where $\alpha_r={\rm Re}\{\alpha\}$, $\alpha_i={\rm Im}\{\alpha\}$, while $s_j = \Omega_j / g_j$, $\Gamma_j = \gamma_j'/2$, with 
$$\gamma_j'=\gamma_j \frac{g_j^{2}}{\Delta_j^{2} + \gamma_j^{2}/4}$$
the effective rate of spontaneous emission. Moreover,
$\kappa' = \kappa + \sum_{j=1}^N \Gamma_j \cos^2(k x_j)$ is the rate at which cavity photons are lost (via both cavity decay and spontaneous emission), and 
$$\Delta_c' = \Delta_c - \sum_{j=1}^N U_j \cos^2(kx_j)$$ is the effective detuning between cavity and laser, which includes the dynamical Stark shift due to the coupling with the atoms. 

Equation \eqref{FPEend} is a Fokker-Planck equation for the variables  $\bm{x}$, $\bm{p}$, $\alpha_r$, and $\alpha_i$. We note that its derivation does not require to explicitly assume a time-scale separation between the cavity field and atomic motion. On the other hand, its derivation consists in neglecting derivatives corresponding to orders $\epsilon_{1,j}\epsilon_{2}^2$, $\epsilon_{1,j}^2\epsilon_{2}$, which is motivated under the assumption that the semiclassical limit for the field amplitude applies. Such approximation becomes invalid for small photon numbers, and thus for instance below and close to the selforganization threshold. Nevertheless, this equation is  used in the literature for studying the dynamics of the system below threshold \cite{Asboth:2005,Griesser}.

\section{Dynamics for low pump intensities}
\label{Sec:numerical}
 
We now analyse the predictions of Eq. \eqref{FPEsemi} for low pump intensities $\Omega$, such that the cavity field is essentially in the vacuum state. We first solve Eq. \eqref{FPEsemi} at the asymptotics of the dynamics and find an explicit form for the stationary distribution and then extract numerical predictions based on stochastic differential equations that we will define below. In the following discussion we will assume for simplicity that all atoms are identical and set $m_j=m$, $S_j=S$, $U_j=U$, and $\gamma_j=\gamma$. 

\subsection{Fokker-Planck equation for small intracavity photon numbers}
\label{Sec:FPEsmall}

We assume that the low effective pumping rate $S$ is small compared with the cavity decay rate $\kappa$ (more precisely, $\sqrt{N}S\ll\kappa$). Consequently, the mean number of intracavity photons is close to zero, $|\alpha|^2\ll 1$. In this limit we can analytically evaluate $\sigma_s$ by reducing the Hilbert space of the photon field to zero and one-photon states. The coefficients can be then analytically determined in lowest order in $|\alpha|$ and Eq. \eqref{FPEsemi} can be cast in the form 
\begin{align}
\label{FPEbelow}
&\frac{\partial}{\partial t} f(\bm{x},\bm{p},t)= - \sum_{n=1}^N  \frac{\partial}{\partial x_n} \frac{p_n}{m} f \\
&-  2 \hbar k S^2\sum_{\ell,n=1}^N \frac{\partial}{\partial p_n}  \frac{\Delta_c'}{\Delta_c'^2 + \kappa^2} \sin(kx_n)\cos(kx_\ell)  f \nonumber \\
&- \frac{4\hbar k^2}{m} S^2\sum_{\ell,n =1}^N \frac{\partial}{\partial p_n} \frac{\Delta_c' \kappa}{(\Delta_c'^2 + \kappa^2)^2} \sin(kx_n) \sin(kx_\ell)  \, p_\ell f  \nonumber \\
&+ \hbar^2 k^2 S^2 \sum_{\ell,n=1}^N \frac{\partial^2}{\partial p_n \partial p_\ell}  \frac{\kappa}{\Delta_c'^2 + \kappa^2} \sin(kx_n) \sin(kx_\ell)  f\,,\nonumber
\end{align}
where we have additionally assumed that $s = \Omega/g = S/U \gtrsim 1 $. 

In Eq. \eqref{FPEbelow} we did not report the cross-derivative between position and momentum, since we assume that the atoms spatial density $n_{\rm at}$ is uniform: In fact, the pump intensity is taken to be well below the selforganization threshold, the mean intracavity photon number is close to zero and we expect that the intracavity optical lattice does not confine the atoms. We will check the consistency of this hypothesis later on. 

We have also neglected spontaneous emission, since we choose $|\Delta|\gg\gamma/2$ and consider large cooperativity, $C = g^2/(\kappa\gamma/2)$ \cite{Kimble}. This can be checked when comparing the contribution to the diffusion coefficient due to spontaneous decay to the one due to the coupling with the cavity field. Their ratio reads
\begin{align*}
\frac{\overline{u^2}}{\langle\sin^2(kx_n)\rangle} \left(\frac{\kappa\gamma}{2g^2}\right) \frac{\Delta_c^{2} + \kappa^2}{\kappa^2} \frac{\Delta^2 + \gamma^2/4}{\Delta^2} 
\cong 2\overline{u^2}\frac{1}{C} \frac{\Delta_c^2 + \kappa^2}{\kappa^2}\,,
\end{align*}
where we have used that $|\Delta_c|\gg N|U|$ and that $\langle \sin^2(kx_j)\rangle=1/2$ when the atoms are not spatially localized inside the volume of the cavity mode. Therefore, the effect of spontaneous decay can be neglected when $C\gg 1$ (and $\Delta_c$ is of the order of $\kappa$), which are the conditions we consider in the following. In the other regime, when the atoms are localized at the points where their coupling with the field is maximum (the selforganized phase), then $\langle \sin^2(kx_j)\rangle\approx 0$ and diffusion is mainly due to spontaneous emission. 

\subsubsection{Stationary state}

We first analyse the predictions of the Fokker-Planck equations under plausible assumptions, which we then verify numerically later on. We extract in particular the cooling rate and steady-state momentum distribution. In the following we assume  $N|U| \ll |\Delta_c|$, which is consistent with uniform spatial distributions, as shown in the following. 

Let us first define the momentum distribution at time $t$, which is the integral of the Wigner function over the positions:
\begin{align*}
F (\bm{p},t) = \int {\rm d}\bm{x} f(\bm{x}, \bm{p}, t)\,.
\end{align*}
Under the assumption of uniform spatial distribution, then $f (\bm{x}, \bm{p},t)\approx F(\bm{p},t) n_{\rm at}$, where we denote the spatial density by $n_{\rm at}$. We then integrate  Eq. \eqref{FPEbelow} over $\bm{x}$ and obtain an equation for the momentum distribution of the form
\begin{align}
\frac{\partial}{\partial t} F(\bm{p},t) 
\approx&-  4 \omega_r\sum_{n=1}^N \frac{\partial}{\partial p_n}  S^2 \frac{\Delta_c \kappa \delta_1}{(\Delta_c^2 + \kappa^2)^2}   \, p_n F(\bm{p},  t) \label{FP:1}\\
&+\hbar m \omega_r \sum_{n=1}^N \frac{\partial^2}{\partial p_n^2} S^2 \frac{\kappa \delta_2}{\Delta_c^2 + \kappa^2}  F(\bm{p},  t)\,,\nonumber
\end{align}
which has been obtained assuming $N|U| \ll |\Delta_c|$, with 
\begin{align*}
\delta_1 &= 1 + \frac{3 \Delta_c^2 - \kappa^2}{\Delta_c^2 + \kappa^2} \frac{NU/2}{\Delta_c} \frac{2N-1}{2N} \,,\\
\delta_2 &= 1+ \frac{2 \Delta_c^2}{\Delta_c^2 + \kappa^2} \frac{NU/2}{\Delta_c} \frac{2N-1}{2N} \,.
\end{align*}
We note that the assumption $N|U| \ll |\Delta_c|$ is consistent with uniform spatial distributions. In Eq. \eqref{FP:1}
there are no terms which mix variables from different atoms: In fact, for uniform spatial distributions they vanish after integrating over the positions. In this limit, the equations for the momentum of each atom can be decoupled using the ansatz $F(\bm{p},t)= \prod_{j=1}^N F_j(p_j,t)$, which delivers the equation of motion for the momentum distribution $ F_j(p_j,t)$ for atom $j$:
\begin{align}
\frac{\partial}{\partial t} F_j(p_j,t) |_t 
=&- \frac{\partial}{\partial p_j}  A p_j F_j(p_j,  t) 
+  \frac{\partial^2}{\partial p_j^2} B F_j(p_j,  t) 
\label{coolone}
\end{align}
with
\begin{align*}
A &= 4 \omega_r S^2 \frac{\Delta_c \kappa \delta_1}{(\Delta_c^2 + \kappa^2)^2}\,, \\
B &= \frac{1}{2} (\hbar k)^2 S^2 \frac{\kappa \delta_2}{\Delta_c^2 + \kappa^2} \,.
\end{align*}
A stationary solution exists for $A<0$, that is verified when $\Delta_c < 0$. In this case, one can make the ansatz that  the momentum distribution of one atom is a Gaussian of width $\Delta p_j$. From Eq. \eqref{coolone} we find that $\Delta p_j=\Delta p(t)$, which is given by the equation
\begin{align}
\label{Delta:p}
\Delta p (t) &= \Bigg\{  \Delta p(0)^2 {\rm e}^{2 A t} +(1- {\rm e}^{2 A t} )\Delta p_{\infty}^2\Bigg\}^{1/2}\,,
\end{align}
with $ \Delta p(0)$ the width at $t=0$ and $\Delta p_{\infty}=\sqrt{-B/A}$ the width of the stationary state. Since the momentum distribution is a Maxwell-Boltzmann distribution, we can associate a temperature $T$ to the width, with
$k_B T = \Delta p_\infty^2/m$, whereby
\beq 
k_BT=\hbar \frac{\Delta_c^2 + \kappa^2}{-4 \Delta_c}\, \frac{\delta_2}{\delta_1}.
\label{Temperature}
\eeq
The steady state is reached with rate $\Gamma_{\rm cool}=-2A$, which we denote by the cooling rate and reads
\begin{align*}
\Gamma_{\rm cool} = 8 \omega_r  S^2 \frac{|\Delta_c| \kappa \delta_1}{(\Delta_c^2 + \kappa^2)^2}\,.
\end{align*}
The cooling rate thus scales with the square of the scattering amplitude $S$ and with the recoil frequency $\omega_r$ and is independent on the number of atoms. In fact, when the atoms spatial distribution is uniform, superradiant effects are negligible and the atoms can be considered as independent scatterers. Minimum temperature and faster rate are found for $\Delta_c=-\kappa$. For this choice, $k_B T =\hbar\kappa/2$ and $\Gamma_{\rm cool}=2\omega_r (S/\kappa)^2 \delta_1$. It is interesting to observe that, also in the case when the cavity is driven, the final temperature of the atomic ensemble is determined by the cavity linewidth, and it is minimal for $\Delta_c \approx - \kappa$ \cite{Vukics:2005}.

\subsubsection{Limits of validity}

On the basis of the results we just derived, we are now able to identify the parameter regime for which the semiclassical description of the atomic motion we applied is valid at the final stages of the cooling dynamics.  
It is simple to check using Eq. \eqref{Temperature} that both conditions \eqref{cond1} and \eqref{cond:2} are verified provided that $|\Delta_c|\sim \kappa$ and that $\omega_r\ll \kappa$. In particular, the requirement that the motion evolves slower than the cavity field, Eq. \eqref{cond:2}, leads to the restriction that the detuning between cavity field and pump cannot be either much larger or much smaller than the cavity linewidth. 

Let us now consider the assumption that the atoms spatial distribution is uniform in space. This assumption shall be checked, since the atoms are subject of the dispersive potential due to the mechanical effects of the cavity field on their motion. Using a uniform spatial distribution in Eq.  \eqref{alpha:x}, one finds that the mean field amplitude vanishes. The mean intracavity  photon number is found using Eq. \eqref{n:cav} for $\mathcal F=a^{\dagger}a$, and reads
\beq
\label{n:cav:1}
n_{\rm cav}\approx \frac{NS^2/2}{\Delta_c^2+\kappa^2}\,,
\eeq
which discards the higher-order contribution due to non-adiabatic effects. This value is much smaller than unity provided that $\sqrt{N} S\ll |\Delta_c+{\rm i}\kappa|$. The corresponding potential depth is $U_0=Un_{\rm cav}$, and it is much smaller than the mean kinetic energy (thus, the atoms are not spatially confined by the potential) when $U_0\ll \kappa/2$, which corresponds to the condition
\beq
\sqrt{N}S\ll \kappa\sqrt{\frac{\kappa}{U}}\,.
\eeq
When $S$ or $N$ are such that this inequality is not fulfilled, the assumption of spatial flat distribution becomes invalid. Correspondingly, the cavity field starts to establish correlations between the atoms which ultimately lead to the formation of ordered structures.

\subsection{Numerical results}

In this section we evaluate the dynamics predicted by the Fokker-Planck equation obtained in the semiclassical limit by adiabatically eliminating the cavity degrees of freedom. Our aim is to get an insight in the dynamics of the system by analysing the trajectories of the atoms. For this purpose, we use Ito-type Stochastic Differential Equations (SDE) \cite{Gardiner:Noise,Gardiner:Stochastic,Horak:JPB}, which we extract from Eq. \eqref{FPEsemi}. They read
\begin{align}
\label{SDE:below}
{\rm d}x_{j} &= \frac{p_{j}}{m} {\rm d}t + {\rm d}X_j,\\
{\rm d}p_{j} &= 2 \hbar k S^2 \sum_{i=1}^N \frac{\Delta_c'}{\Delta_c'^2 + \kappa^2} \cos(k x_i) \sin(k x_j) {\rm d}t \label{SDE:below2} \\
&+ \sum_{i=1}^N 8 \omega_r S^2 \frac{\Delta_c' \kappa}{(\Delta_c'^2 + \kappa^2)^2} \sin(kx_i) \sin(kx_j) p_i {\rm d}t + {\rm d}P_{j}, \nonumber
\end{align}
where $j = 1,...,N$ labels the atoms and ${\rm d}P_j$ denotes the noise term, which is simulated by means of a Wiener process. In particular, $\langle {\rm d}P_j\rangle=0$ and $\mean{{\rm d}P_i {\rm d}P_j}= 2 D_{ij} dt$ with 
\begin{align}
D_{ij} = (\hbar k)^2 S^2 \frac{\kappa}{\Delta_c'^2 + \kappa^2} \sin(k x_i) \sin(k x_j)
\label{Dcav}
\end{align}
the element of the diffusion matrix, while $\mean{{\rm d}P_j {\rm d}X_\ell} =  \eta_{j\ell} dt$, with
\begin{align}
\eta_{j\ell} = 2 \hbar \omega_r S^2 \sin(k x_{j}) \sin(k x_{\ell})  \frac{\kappa^2 - \Delta_c'^2}{(\Delta_c'^2 + \kappa^2)^2}\,.
\label{eta}
\end{align}

When one includes spontaneous emission, the elements of the diffusion matrix read
\begin{align}
D_{ij} = (\hbar k)^2 \big\{ \frac{S^2 \kappa}{\Delta_c'^2 + \kappa^2} \sin(k x_i) \sin(k x_j) + \delta_{ij} \frac{\gamma'}{2} s^2 \overline{u^2} \big\},
\label{Dspon}
\end{align}
which reports the dominant contributions (the rescattering of a cavity photon by the atom is here neglected, see Eq. \eqref{D'}).
The analytical estimate of the steady-state momentum width for homogeneous spatial distribution increases accordingly
\begin{align}
\Delta p_{\infty}=\sqrt{-B/A} \big\{ 1 + 2 \overline{u^2} \frac{\Delta^2 + \gamma^2/4}{\Delta^2} \frac{\Delta_c^2 + \kappa^2}{\kappa^2 C \delta_2} \big\}^{1/2}.
\label{dpspon}
\end{align}

The simulations are performed considering a gas of $^{85}\text{Rb}$ atoms, whose $\text{D}_2$-line, namely, the   hyperfine transition $5^{2}\text{S}_{1/2} \leftrightarrow 5^{2}\text{P}_{3/2}$ at wave length $\lambda = 780$ nm and linewidth $\gamma /2 = 2 \pi \times 3 $ MHz, couples with the mode of the resonator and with the transverse laser. 
The laser frequency is assumed to be detuned from the atomic frequency by $\Delta_a = - 500 \gamma /2$, and from the cavity frequency by $\Delta_c$, with $N$ the number of atoms and $\kappa=0.5 \gamma/2$. The dynamics and steady state of the atoms is studied, assuming that initially the atoms are at steady state of Doppler cooling with $k_B T = \hbar \gamma /2$. The initial state is a Gaussian distribution (the initial momentum is generated by means of Gaussian-distributed random numbers) with uniform density (the initial positions of the atoms are given by means of uniformly distributed random numbers in the interval $[0, \lambda]$).  

\begin{figure}[htbp]
\centering
\includegraphics[width=0.48\textwidth]
{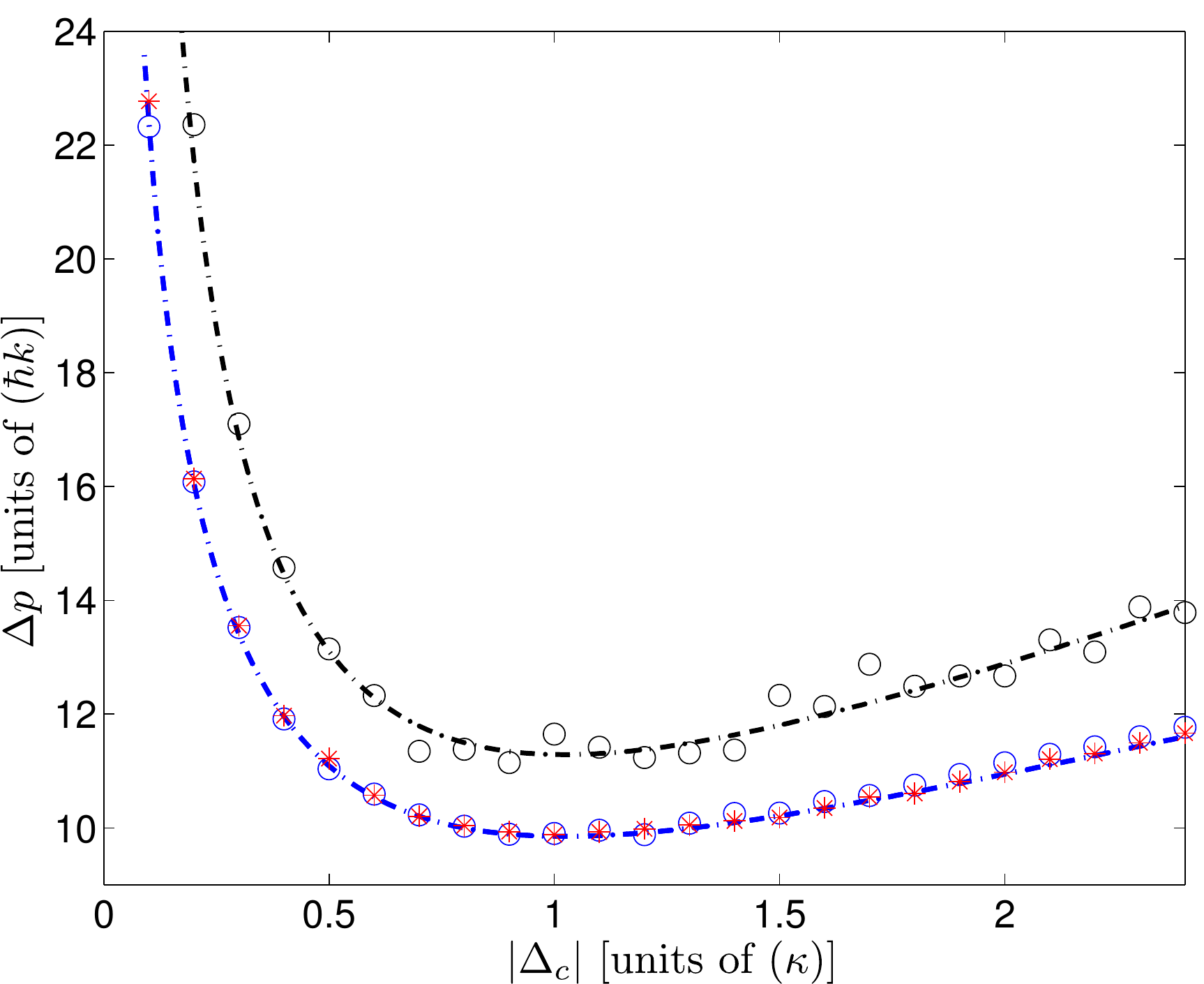}
\caption{(Color online) Width of the steady-state momentum distribution as a function of $\Delta_c$. The blue circles correspond to the result of numerically integrating 5000 trajectories using Eqs. \eqref{SDE:below}-\eqref{SDE:below2} for each value of $\Delta_c$ together with Eq. \eqref{Dcav}. The blue dashed line plots Eq. \eqref{Delta:p}. The red stars correspond to the result of numerically integrating 5000 trajectories corresponding to the Fokker-Planck equation \eqref{FPEend}, where the cavity field is treated in the semiclassical limit (details are reported in Appendix B). The results are obtained by neglecting spontaneous emission.  Spontaneous emission is included in the results reported by the black circles, which have been obtained by integrating Eqs. \eqref{SDE:below}-\eqref{SDE:below2} with Eq. \eqref{Dspon} for  200 trajectories. The black dashed line gives the corresponding analytical estimate according to Eq. \eqref{dpspon}. The deviations of the simulation data from the fit originate from statistical noise. The parameters are: $\kappa = 0.5 \gamma/2$, $NU / \Delta_c = 0.05$, $\Omega/\Omega_c = 0.3$, $N=5$. 
\label{fig:steady}}
\end{figure}

In the calculations we neglect spontaneous emission, which is plausible under the assumption that the cavity is far-off resonance from the atomic transition. We have checked for a sample of values when this assumption is justified by comparing the simulations including spontaneous emission with the simulations in which spontaneous decay was not included (see Figure \ref{fig:steady} and related discussion). We further discard the cross-correlations, setting  $\mean{{\rm d}P_j {\rm d}X_\ell} =  0$, after verifying that this assumption is justified for the considered parameter choice \cite{footnote}.  We first check that the parameters are chosen, so that the number of intracavity photons is sufficiently close to zero. We choose $\Delta_c = - \kappa$, for which one expects the minimum value from Eq. \eqref{Temperature}. Using Eq. \eqref{n:cav} and setting $n_{\rm cav}=0.1$, we obtain $\sqrt{N}\Omega\sim 0.6 |\Delta_a|\kappa/g$. For the parameters we chose and $NU/\Delta_c=0.05$, $\Omega\sim 45 \gamma/2$. We set $\Omega \sim 21 \gamma/2$, which corresponds to $n_{\rm cav} \sim 0.02$. In this regime, we evaluate the density matrix of the field in the reduced Hilbert space, where the photon states are truncated till $n=2$, see Sec. \ref{Sec:FPEsmall}.

\begin{figure}[htbp]
\centering
\includegraphics[width=0.48\textwidth]
{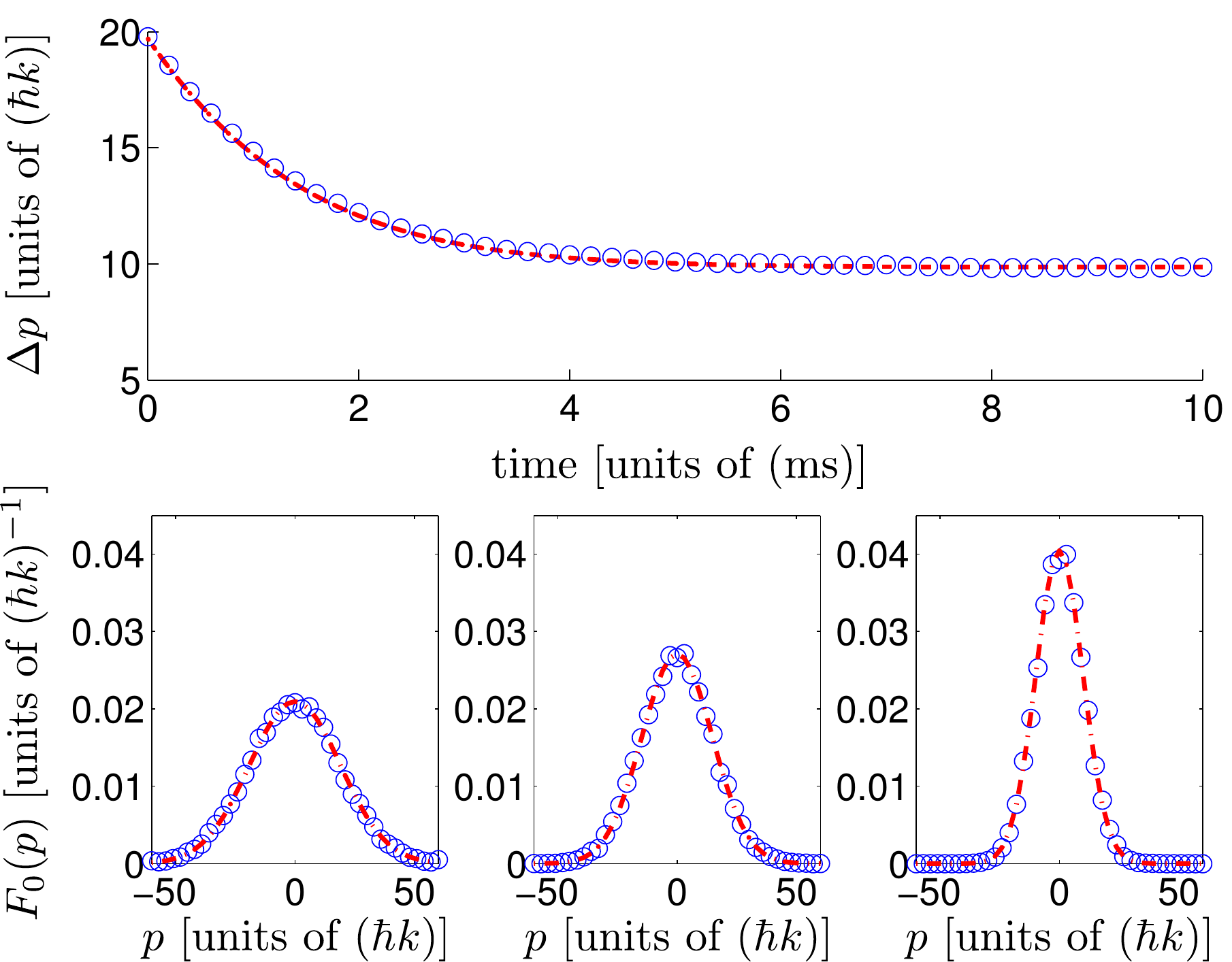}
\caption{(Color online) Time evolution of the momentum distribution $F_0(p,t)$ evaluated from 5000 trajectories simulated by integrating Eqs. \eqref{SDE:below}-\eqref{SDE:below2} for $N=5$ atoms. The top panel gives $\Delta p(t)$ in units of the momentum recoil $\hbar k$ (circles). The dashed line corresponds to Eq. \eqref{Delta:p}. Lower panels: Momentum distribution at $t=(0.1,1,9)$ ms (circles) compared to a Gaussian of width $\Delta p(t)$ given by Eq.  \eqref{Delta:p} (dashed). The parameters are: $\kappa = 0.5 \gamma/2$, $NU / \Delta_c = 0.05$, $\Omega \sim 21 \gamma/2$, $\Delta_c = - \kappa$. The  initial momentum distribution is a Gaussian with $k_BT_{\rm in}= \hbar \gamma / 2$ for each atom.\label{fig:time}}
\end{figure}

On the basis of this result, we evaluate the time evolution of the width $\Delta p$ of the momentum distribution for each atom taking $N=5$ atoms.
If we take identical particles with the same initial temperature, the momentum distribution of each atom will be the same at all times. We then focus on the momentum distribution averaged over all the atoms
\begin{align*}
F_0 (p) = \frac{1}{N} \sum_{j=1}^{N} \int_{-\infty}^{\infty} {\rm d}p_j \delta (p-p_j) F_j(p_j).
\end{align*} 
Figure \ref{fig:time} displays the width of the momentum distribution as a function of time: The points are obtained from 5000 trajectories for an initial momentum distribution corresponding to a Maxwell-Boltzmann distribution with $k_B T = \hbar \gamma / 2$ for each atom. The dashed line is the function given in Eq. \eqref{Delta:p}, which has been obtained by assuming that all atoms are independently cooled and show excellent agreement with the numerics. The lower panels show the momentum distribution at given instants of times, $t=(0.1,1,9)$ ms. The dashed line corresponds to the prediction extracted from Eq. \eqref{coolone}, which gives a Gaussian at all times.

\begin{figure}[htbp]
\centering
\includegraphics[width=0.48\textwidth]
{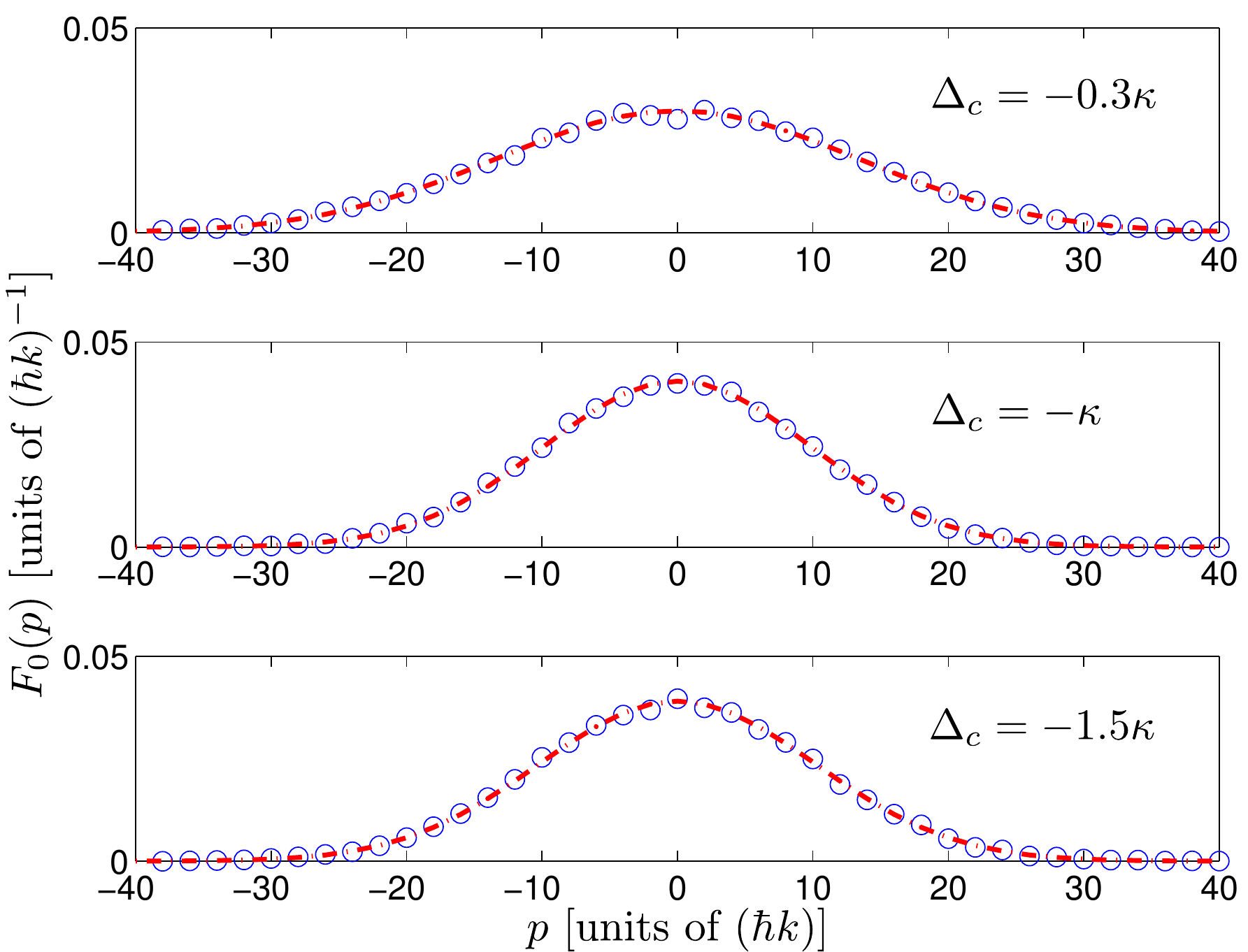}
\caption{(Color online) Steady state momentum distribution (circles) for $\Delta_c=-0.3 \kappa$ (top), $\Delta_c=-\kappa$ (middle) and $\Delta_c=-1.5 \kappa$ (bottom). The points are extracted from 5000 trajectories evaluated for each value $\Delta_c$ by integrating the SDE in Eqs. \eqref{SDE:below}-\eqref{SDE:below2} . The dashed line is a Gaussian whose width is given by Eq. \eqref{Delta:p}. The parameters are: $\kappa = 0.5 \gamma/2$, $NU / \Delta_c = 0.05$, $\Omega/\Omega_c = 0.3$, $N=5$.  The initial momentum distribution is a Maxwell Boltzmann  with mean kinetic energy $k_B T_{\rm in} = \hbar \gamma / 2$ for each atom.}
\label{fig:steady3}
\end{figure}

We now analyse the dependence of the final temperature on the detuning $\Delta_c$. In order to perform a systematic comparison with the predictions of the Fokker-Planck equation in Ref. \cite{Griesser}, we express the 
pumping strength $\Omega$ of the laser in units of the self-organization threshold defined as \cite{Griesser}
\begin{align*}
|\Omega_c| = \frac{\kappa^2+\delta^2}{2|\delta|\sqrt{N}}\frac{|\Delta_a|}{g}\,,
\end{align*} 
with $\delta=\Delta_c-NU/2$. This value scales with the number of atoms and the detuning $\Delta_c$. We fix $\Omega=0.3\,\Omega_c$, for which the mean photon number, Eq. \eqref{n:cav}, takes the form
\begin{align*}
n_{\rm cav} \approx \left(\frac{\Omega}{\Omega_c}\right)^2 \frac{\Delta_c^2+ \kappa^2}{8 \Delta_c^2} .
\end{align*}
This equation shows that, when $|\Delta_c|$ becomes too small, the number of intracavity photons increases like $n_{\rm cav} \propto \frac{1}{\Delta_c^2}$.\\
Figure \ref{fig:steady3} displays the momentum distribution at the asymptotics of the dynamics, which is found by integrating the SDE after several ms, for three values of $\Delta_c$. The curves are fitted by Gaussian of width  given in Eq. \eqref{Delta:p}: The stationary momentum distribution is thus a Gaussian with width $\Delta p_{\infty}=\sqrt{m k_BT}$ with $T$ given in Eq. \eqref{Temperature}.

Figure \ref{fig:steady} shows the stationary momentum width as a function of $\Delta_c$, which has been extracted by numerically integrating the SDE (see blue circles). The curve exhibits a minimum at $\Delta_c=-\kappa$ and is in excellent agreement with Eq. \eqref{Delta:p}, evaluated at $t\to\infty$ at the corresponding value of $\Delta_c$ (see blue dashed line). We have compared the predictions of the Fokker-Planck equation, Eq. \eqref{FPEsemi}, with the ones of the Fokker-Planck equation in Eq. \eqref{FPEend}, based on the assumption that the cavity field can be treated semiclassically. The simulations are performed by integrating the SDE reported in Ref. \cite{Horak:JPB}, which for completeness are reported in Appendix B (see red stars). Agreement between the predictions of the two Fokker-Planck equations is found: This is remarkable, since the cavity field in this regime is in the vacuum, and thus outside the formal limits of validity of Eq. \eqref{FPEend}. In order to check the effect of spontaneous emission, we have integrated Eqs. \eqref{SDE:below}-\eqref{SDE:below2} using Eq. \eqref{Dspon}.  The result is shown by the black circles. The black dashed line corresponds to $\Delta p_{\infty}$ in Eq. \eqref{dpspon} and fits the numerical data. We observe an increase of $\Delta p_{\infty}$ by about 15 $\%$ with respect to the case in which spontaneous emission is not included. 

We finally comment on the type of momentum distribution we find. In \cite{Griesser}, by studying the equation of motion of atoms inside a resonator and driven well below threshold, it was argued that the steady state momentum distribution for the atoms far below threshold can obey a q-Gaussian distribution with $q=1+ \omega_r / |\delta|$, where $\delta = \Delta_c - NU/2$ and $N|U| \ll |\Delta_c|$. This calculation was performed by neglecting spontaneous decay. In our model, we do find Gaussian distributions ($q \approx 1$), whose steady-state temperature, Eq. \eqref{Temperature}, is comparable with the result in \cite{Griesser}. Indeed, our model can only allow for regular Gaussian distribution. In fact, this is consistent with the limits of validity of the  Fokker-Planck equation we derive, which requires the separation of the time scales between cavity field and atoms dynamics, namely, $|\Delta_c| \approx \kappa \gg \omega_r$. 

\section{Discussion and Conclusions}
\label{Sec:Conclusions}

In this article we have derived a Fokker-Planck equation which describes the dynamics of atoms which are cooled by radiative processes, where laser photons are scattered into the mode of a high-finesse resonator. The derivation is based on the assumption that the time scale of the atomic center-of-mass motion dynamics is much larger than the cavity field typical time scales, and thus holds for resonators whose linewidth $\kappa$ is much larger than the atomic recoil frequency $\omega_r$. It cannot be applied, thus, to the setup in Ref. \cite{Wolke:2012}. In this limit, $\kappa \gg \omega_r$, we eliminate the cavity field from the atomic motion dynamics using a perturbative expansion up to second order in the retardation effects and derive a Fokker-Planck equation for the atomic variables. The equation we derive constitutes an alternative theoretical description for the cavity cooling dynamics of atomic ensembles in high-finesse resonators. In particular, our model provides a description which is not restricted by the value of the laser intensity. 

We have analysed the predictions when the laser intensity is well below the selforganization threshold. In this limit collective effects can be discarded and there is no spatial localization of the atoms by the light forces: at the steady state of the dynamics the atoms  are uniformly distributed inside the cavity and their momenta obey a Maxwell Boltzmann distribution, whose width is determined by the cavity linewidth \cite{Hechenblaikner:1998}. This result is in agreement with previous studies based on other approaches \cite{Asboth:2005,Griesser}. In future studies we will apply this formalism to the dynamics of atoms and field when the laser intensity is close to threshold. This will allow us to investigate the onset of selforganization, and to predict, amongst others, the coherence properties of the light emitted by the resonator.

\acknowledgements
The authors are grateful to A. Vukics, W. Niedenzu and H. Ritsch for stimulating discussions and helpful comments. This work was supported by the European Commission (IP AQUTE, STREP PICC), the BMBF (QuORep), and the German Research Foundation.

\begin{appendix}

\section{Spontaneous emission}

In this appendix we report the Lindbladian in Eq. \eqref{Wigner:Below} and the coefficients of Fokker-Planck Eq. \eqref{FPEsemi} which are due to spontaneous decay. For simplicity, we assume $k_j = k$. 
\begin{widetext}
The Lindbladian in Eq. \eqref{Wigner:Below} reads
\begin{eqnarray}
 &&\mathcal{L}^{\gamma} \hat{\tilde{W}}_t \begin{pmatrix} \bm{x} \\ \bm{p}  \end{pmatrix} \nonumber
 \\
 = 
 &&- \sum_j \frac{\gamma_j'}{2} s_j^2 \Bigg\{ 2 \hat{\tilde{W}}_t \begin{pmatrix} \bm{x} \\ \bm{p}  \end{pmatrix}  - 2 \int_{-1}^1 {\rm d}u N_j (u) \hat{\tilde{W}}_t \begin{pmatrix} \bm{x} - \frac{\tau}{m_j} \hbar \bm{k_j} u \\ \bm{p} + \hbar \bm{k_j} u  \end{pmatrix} \Bigg\}  \\
&&- \sum_j \frac{\gamma_j'}{8} \Bigg\{ \hat{a}^{\dagger} \hat{a} \bigg[ 
{\rm e}^{2 {\rm i} k (x_j + \frac{p_j}{m_j} \tau)}  \hat{\tilde{W}}_t \begin{pmatrix} \bm{x} + \frac{\tau}{m_j} \hbar \bm{k}_j \\ \bm{p} - \hbar \bm{k_j}  \end{pmatrix}
+ {\rm e}^{-2 {\rm i} k (x_j + \frac{p_j}{m_j} \tau)}  \hat{\tilde{W}}_t \begin{pmatrix} \bm{x} - \frac{\tau}{m_j} \hbar \bm{k}_j \\ \bm{p} + \hbar \bm{k_j}  \end{pmatrix} \bigg]
\nonumber \\
&&\qquad\qquad\quad\,\, + \bigg[
{\rm e}^{2 {\rm i} k (x_j + \frac{p_j}{m_j} \tau)} \hat{\tilde{W}}_t \begin{pmatrix} \bm{x} - \frac{\tau}{m_j} \hbar \bm{k}_j \\ \bm{p} + \hbar \bm{k_j}  \end{pmatrix} 
+ {\rm e}^{-2 {\rm i} k (x_j + \frac{p_j}{m_j} \tau)} \hat{\tilde{W}}_t \begin{pmatrix} \bm{x} + \frac{\tau}{m_j} \hbar \bm{k}_j \\ \bm{p} - \hbar \bm{k_j}  \end{pmatrix}  \bigg] \hat{a}^{\dagger} \hat{a} 
+ 2 \big\{ \hat{a}^{\dagger} \hat{a}, \hat{\tilde{W}}_t \begin{pmatrix} \bm{x} \\ \bm{p}  \end{pmatrix} \big\}_+ \nonumber \\
&&\qquad\qquad -2 \int_{-1}^{1} N_j(u) 
\hat{a} \bigg[   {\rm e}^{2 {\rm i} k (x_j + \frac{p_j}{m_j} \tau)}  \hat{\tilde{W}}_t \begin{pmatrix} \bm{x} - \frac{\tau}{m_j} \hbar \bm{k}_j u \\ \bm{p} + \hbar \bm{k_j} u  \end{pmatrix}   
+  \hat{\tilde{W}}_t \begin{pmatrix} \bm{x} - \frac{\tau}{m_j} \hbar \bm{k}_j (u-1) \\ \bm{p} + \hbar \bm{k_j} (u-1)  \end{pmatrix}  \nonumber \\
&&\qquad\qquad\qquad\qquad\qquad
+ \hat{\tilde{W}}_t \begin{pmatrix} \bm{x} - \frac{\tau}{m_j} \hbar \bm{k}_j (u+1) \\ \bm{p} + \hbar \bm{k_j} (u+1)  \end{pmatrix}  
+ {\rm e}^{-2 {\rm i} k (x_j + \frac{p_j}{m_j} \tau)} \hat{\tilde{W}}_t \begin{pmatrix} \bm{x} - \frac{\tau}{m_j} \hbar \bm{k}_j u \\ \bm{p} + \hbar \bm{k_j} u  \end{pmatrix}  \bigg]  \hat{a}^{\dagger} {\rm d}u \Bigg\} \nonumber \\
&&- \sum_j \frac{\gamma_j'}{2} s_j \Bigg\{ (\hat{a} + \hat{a}^{\dagger}) \frac{1}{2} \bigg[ 
{\rm e}^{{\rm i} k (x_j + \frac{p_j}{m_j} \tau)} \hat{\tilde{W}}_t \begin{pmatrix} \bm{x} + \frac{\tau}{2m_j} \hbar \bm{k_j} \\ \bm{p} - \frac{1}{2} \hbar \bm{k_j}  \end{pmatrix}
+ {\rm e}^{-{\rm i} k (x_j + \frac{p_j}{m_j} \tau)} \hat{\tilde{W}}_t \begin{pmatrix} \bm{x} - \frac{\tau}{2m_j} \hbar \bm{k_j} \\ \bm{p} + \frac{1}{2} \hbar \bm{k_j}  \end{pmatrix} \bigg] \nonumber \\
&&\qquad\qquad\qquad\qquad + \frac{1}{2} \bigg[ {\rm e}^{{\rm i} k (x_j + \frac{p_j}{m_j} \tau)} \hat{\tilde{W}}_t \begin{pmatrix} \bm{x} - \frac{\tau}{2m_j} \hbar \bm{k_j} \\ \bm{p} + \frac{1}{2} \hbar \bm{k_j}  \end{pmatrix} 
+{\rm e}^{-{\rm i} k (x_j + \frac{p_j}{m_j} \tau)} \hat{\tilde{W}}_t \begin{pmatrix} \bm{x} + \frac{\tau}{2m_j} \hbar \bm{k_j} \\ \bm{p} - \frac{1}{2} \hbar \bm{k_j}  \end{pmatrix} \bigg] (\hat{a} + \hat{a}^{\dagger}) \nonumber \\
&&\qquad\qquad- \int_{-1}^{1} N_j (u) \bigg( 
\hat{a} \bigg[ {\rm e}^{{\rm i} k (x_j + \frac{p_j}{m_j} \tau)}  \hat{\tilde{W}}_t \begin{pmatrix} \bm{x} - \frac{\tau}{m_j} \hbar \bm{k_j} (u-\frac{1}{2}) \\ \bm{p} + \hbar \bm{k_j} (u - \frac{1}{2})  \end{pmatrix}
+ {\rm e}^{-{\rm i} k (x_j + \frac{p_j}{m_j} \tau)} \hat{\tilde{W}}_t \begin{pmatrix} \bm{x} - \frac{\tau}{m_j} \hbar \bm{k_j} (u+\frac{1}{2}) \\ \bm{p} + \hbar \bm{k_j} (u + \frac{1}{2})  \end{pmatrix} \bigg] \nonumber \\
&&\qquad\qquad\qquad\qquad\qquad + \bigg[ {\rm e}^{{\rm i} k (x_j + \frac{p_j}{m_j} \tau)} \hat{\tilde{W}}_t \begin{pmatrix} \bm{x} - \frac{\tau}{m_j} \hbar \bm{k_j} (u+\frac{1}{2}) \\ \bm{p} + \hbar \bm{k_j} (u + \frac{1}{2})  \end{pmatrix}
+  {\rm e}^{-{\rm i} k (x_j + \frac{p_j}{m_j} \tau)} \hat{\tilde{W}}_t \begin{pmatrix} \bm{x} - \frac{\tau}{m_j} \hbar \bm{k_j} (u-\frac{1}{2}) \\ \bm{p} + \hbar \bm{k_j} (u - \frac{1}{2})  \end{pmatrix} \bigg] \hat{a}^{\dagger} \bigg)  {\rm d}u   \Bigg\} \,.\nonumber 
\end{eqnarray}
This superoperator is expanded in power of $\epsilon$, such that the zeroth order term in Eq. \eqref{L0:gamma} reads
\begin{eqnarray}
\mathcal{L}_0^{\gamma'} \hat{\tilde{W}} =
 \sum_j \frac{\gamma_j '}{2} s_j \commutator{(\OPa - \OPadagger)}{\cos (kx_j) \hat{\tilde{W}}  }   
-\sum_j \frac{\gamma_j'}{2} \cos^2(k x_j) \big\{ \OPadagger \OPa \hat{\tilde{W}} + \hat{\tilde{W}} \OPadagger \OPa - 2 \OPa \hat{\tilde{W}} \OPadagger \Big\} \,,
\end{eqnarray}
while the first and second order terms, in Eqs. \eqref{L1:gamma} and \eqref{L2:gamma} respectively, take the form
\begin{eqnarray} 
\mathcal{L}_1^{\gamma'}\hat{\tilde{W}} &=& 
- \sum_j \frac{\gamma_j'}{2} (\hbar k) \frac{\partial}{\partial p_j} \frac{- {\rm i}}{2} \sin(2 k x_j) \commutator{\hat{a}^{\dagger} \hat{a}}{\hat{\tilde{W}}} 
- \sum_j \frac{\gamma_j'}{2} (\hbar k) \frac{1}{2} \frac{\tau {\rm i}}{m_j} \sin(2 k x_j) \commutator{\hat{a}^{\dagger} \hat{a}}{\frac{\partial}{\partial x_j} \hat{\tilde{W}}} 
\\
&+&  \sum_j \frac{\gamma_j'}{2} \frac{k p_j}{m_j} \tau \sin(2 k x_j) \big( \hat{a}^{\dagger} \hat{a}  \hat{\tilde{W}} +  \hat{\tilde{W}} \hat{a}^{\dagger} \hat{a} -2 \hat{a}  \hat{\tilde{W}} \hat{a}^{\dagger} \big),   \nonumber \\
\mathcal{L}_2^{\gamma'}\hat{\tilde{W}} 
&=& 
\sum_j \frac{\gamma_j'}{2} s_j^2 (\hbar k)^2 (\overline{u^2})_j \frac{\partial^2}{\partial p_j^2} \hat{\tilde{W}} 
\\
&-& \sum_j \frac{\gamma_j'}{2} \frac{1}{8} (\hbar k)^2 \big\{ 2 \cos(2 k x_j) ( \hat{a}^{\dagger} \hat{a} \frac{\partial^2}{\partial p_j^2} \hat{\tilde{W}} +  \frac{\partial^2}{\partial p_j^2} \hat{\tilde{W}} \hat{a}^{\dagger} \hat{a}    )  
-4 \hat{a} \frac{\partial^2}{\partial p_j^2} \hat{\tilde{W}} \hat{a}^{\dagger} - 2 (\overline{u^2})_j ( 2 \cos(2kx_j) + 2 ) \hat{a} \frac{\partial^2}{\partial p_j^2} \hat{\tilde{W}} \hat{a}^{\dagger} \big\} \nonumber \\
&+& \sum_j \frac{\gamma_j'}{2} s_j (\hbar k)^2      
(\overline{u^2})_j \cos(k x_j) ( \hat{a}  \frac{\partial^2}{\partial p_j^2} \hat{\tilde{W}} +  \frac{\partial^2}{\partial p_j^2} \hat{\tilde{W}} \hat{a}^{\dagger}  ).      \nonumber
\end{eqnarray}
Finally, the coefficients appearing in Eq. \eqref{FPEsemi} and due to spontaneous emission are given by the expressions \\
\\
\begin{eqnarray}
\gamma_{j\ell}' &=& \frac{\gamma_\ell'}{2} \frac{k}{m_{\ell}} \sin(2k x_\ell) \text{Tr} \left\{ \OP{F}_j \int_{ 0}^{\infty} {\rm d}\tau  \exp\left(\mathcal{L}_0 \tau\right) \tau \big( \hat{a}^{\dagger} \hat{a}  \sigma_s(\bm{x}) + \sigma_s(\bm{x}) \hat{a}^{\dagger} \hat{a} -2 \hat{a} \sigma_s(\bm{x}) \hat{a}^{\dagger} \big) \right\}, \\
D_{j\ell}' &=&  -\frac{\gamma_\ell'}{2} (\hbar k) \frac{ {\rm i} }{2 } \sin(2 k x_\ell) \text{Tr} \left\{ \OP{F}_j \int_{ 0}^{\infty} {\rm d}\tau  \exp\left(\mathcal{L}_0 \tau\right)  \commutator{\hat{a}^{\dagger} \hat{a} }{ \sigma_s (\bm{x})} \right\} \label{D'} \\
 &&+ \delta_{j\ell} (\hbar k)^2 \frac{\gamma_j'}{2}\bigg[ \mean{\hat{a}^{\dagger} \hat{a}}_{\sigma_s(\bm{x})}  [ \sin^2(k x_j) +  (\overline{u^2})_j \cos^2( kx_j) ]  + s_j  (\overline{u^2})_j (\mean{  \hat{a} + \hat{a}^{\dagger}  }_{\sigma_s (\bm{x})} \cos(k x_j) +s_j)  \bigg],    \nonumber \\
\eta_{j\ell}' &=& \frac{\gamma_\ell'}{2} \frac{- {\rm i} \hbar k}{2 m_{\ell}} \sin(2 k x_\ell) \text{Tr} \left\{ \OP{F}_j \int_{ 0}^{\infty} {\rm d}\tau  \exp\left(\mathcal{L}_0 \tau\right) \tau \commutator{\hat{a}^{\dagger} \hat{a} }{ \sigma_s (\bm{x})} \right\}.
\end{eqnarray}

\end{widetext}

\section{Fokker-Planck Equation for large photon numbers}

For the Fokker-Planck equation \eqref{FPEend} the SDE take the form \cite{Horak:JPB}
\begin{align}
{\rm d}x_{j} &= \frac{p_{j}}{m_{j}} {\rm d}t,
\label{SDEx} \\
{\rm d}p_{j} &= - \hbar \nabla_j \big[ U_j |\alpha|^2  \cos^2(k x_{j})  + s_j U_j \cos(k x_{j}) (2 \alpha_r)  \nonumber \\ 
&\qquad \quad -  (2 \alpha_i) \Gamma_{j} s_j \cos(k x_{j}) \big] {\rm d}t + {\rm d}P_{j},
\label{SDEp}\\
{\rm d}{\alpha}_r &= \big( - \Delta_c' \alpha_i - \kappa' \alpha_r - \sum_j \Gamma_j s_j \cos(k x_j) \big) {\rm d}t + {\rm d}A_r,
\label{SDEar}\\
{\rm d}{\alpha}_i &= \big( \Delta_c' \alpha_r - \kappa' \alpha_i - \sum_j s_j U_j \cos(k x_j) \big) {\rm d}t + {\rm d}A_i,
\label{SDEai}
\end{align}
where the noise terms $dP_j, dA_r$ and $dA_i$ are simulated by means of Wiener processes 
\begin{align}
\left
(\begin{array}{c} {\rm d}A_r \\ {\rm d}A_i \\ {\rm d}P_1 \\ ... \\ {\rm d}P_N \end{array}\right)  
= B \left(\begin{array}{c} {\rm d}W_1 \\ {\rm d}W_2 \\ {\rm d}W_3 \\ ... \\ {\rm d}W_{N+2} \end{array}\right),
\end{align}   
where $BB^T = D'$. The diffusion matrix now reads \eqref{FPEend} 
\begin{align*}
D' =
\begin{pmatrix}  a            & 0             & -b_1 \alpha_i  &-b_2 \alpha_i & \cdots & -b_N \alpha_i \\
\\
                 0            & a             & b_1 \alpha_r   & b_2 \alpha_r & \cdots &  b_N \alpha_r\\
\\
                -b_1 \alpha_i & b_1 \alpha_r  & c_1            & 0  & \cdots & 0 \\
\\
                -b_2 \alpha_i & b_2 \alpha_r  & 0               & c_2  & \ddots & \vdots \\
\\
                \vdots & \vdots & \vdots & \ddots & \ddots & 0    \\
\\               
                -b_N \alpha_i & b_N \alpha_r & 0 & \cdots & 0 & c_N            \end{pmatrix}
\end{align*}
with $a = \kappa'/2$, $b_j = - (\hbar k/2) \Gamma_j \sin(2 k x_j)$ and
\begin{align*}
c_j &=  2 (\hbar k)^2\Gamma_j  \bigg[  |\alpha|^2  [ \sin^2(kx)  + (\overline{u^2})_j \cos^2(k x_j) ] \nonumber \\ &\qquad \quad + s_j (\overline{u^2})_j (2 \alpha_r \cos(k x_j) +s_j) \bigg].
\end{align*}
When we integrate these stochastic differential equations, we assume that the initial state of the cavity field is a coherent state with $\mean{\alpha_r} = 5$ and $\mean{\alpha_i}=0$. This ensures the validity of the semiclassical description for the cavity field at $t=0$ which has to be verified for all later times.

\end{appendix}

 \singlespacing 


\bibliography{literature}  

\begin{thebibliography}{99}
\bibitem{footnote}
We have performed numerical simulation with and without Eq. \eqref{eta} and found no relevant difference in dynamics and asymptotic behaviour both with and without spontaneous decay.
\end{thebibliography}
\bibliographystyle{unsrt} 

\end{document}